\documentstyle[amstex,amssymb,prb,preprint,aps]{revtex}

\tightenlines
\begin{document}
\title{Non-homogeneous magnetic field induced magnetic edge states and their
transport in a quantum wire}
\author{S. M. Badalyan\cite{smb} and F. M. Peeters\cite{fmp}}
\address{Department of Physics, University of Antwerp (UIA), B-2610 Antwerpen, Belgium}
\date{\today}
\draft
\maketitle

\begin{abstract}
The spectrum of magnetic edge states and their transport properties in the
presence of a perpendicular non-homogeneous magnetic field in a quantum wire
formed by a parabolic confining potential are obtained. Systems are studied
where the magnetic field exhibits a discontinuous jump in the transverse
direction and changes its sign, strength, and both sign and strength at the
magnetic interface. The energy spectra and wave functions of these systems,
the corresponding group velocities along the interface and the particle
average positions normal to the interface are calculated. The resistance of
the quantum wire in the presence of such a magnetic interface is obtained
both in the ballistic and the diffusive regimes as a function of the Fermi
energy and of the homogeneous background magnetic field. The results are
compared with those for the case of a homogeneous field.
\end{abstract}


\section{Introduction}

\noindent Investigations of reduced dimensionality semiconductor systems
is frequently connected with the use of a magnetic field, which, in addition
to the lateral confinement, quantizes the carrier motion also in the plane
normal to the magnetic field. Particularly, a two-dimensional electron gas
(2DEG) exposed to a homogeneous magnetic field has proven to be an extremely
rich subject for investigations in theory and experiment \cite{landwehr92}.

In the last several years a more complex situation of reduced
dimensionality semiconductor systems{\it \ }exposed to a {\it %
non-homogeneous }magnetic field has attracted considerable interest \cite
{peetershb}. Different experimental groups have succeeded in realizing such
systems \cite{ye1,izawa,carmona}. High mobility 2DEGs are formed in standard
GaAs/AlGaAs heterojunctions and the spatial modulation of the magnetic field
is achieved by depositing patterned gates of superconducting or
ferromagnetic materials on top of the heterostructure. An alternative
approach to produce non-homogeneous magnetic fields is by varying the {\it %
topography} of an electron gas \cite{foden1}. These new technologies opened
up a new dimension for investigations of reduced dimensionality
semiconductor systems. Characterizing and understanding transport properties
of these systems are crucial both for fundamental physics and for device
applications.

Theoretically the transport properties of reduced dimensionality
semiconductor systems subjected to a spatial dependent magnetic field have
been addressed in several recent works. The possibilities of the creation of
periodic superstructures by a non-homogeneous magnetic field were
investigated in Refs.~\onlinecite{dubrovin,vilms,peetvasil}.
The magnetic field
dependence of the conductance of a ballistic quantum wire a finite section
of which is subjected to a magnetic field \cite{yoshnag} and of a 2DEG
through an orifice \cite{avishai} was investigated. The single-particle
energy spectrum of a 2DEG subjected to a non-homogeneous magnetic
field was calculated for different step-like \cite{peetmat}, linearly \cite
{muller,hofstetter}, and parabolically (in the transverse direction of a one
dimensional channel) \cite{takagaki} varying with position, and for other
functional magnetic field profiles \cite{calvo,peetmatvas}. It has been 
shown that the
spectrum consists of states that propagate normal to the field gradient and
have remarkable time-reversal asymmetry \cite{muller} in a linearly varying
magnetic field while the spatial distribution of electron and current
densities has a rich structure related to the energy quantization \cite
{hofstetter}. Transport properties of a 2DEG in a magnetic superlattice have
been investigated in weakly \cite{peetvasil} and strongly \cite
{ibrahim2}\ modulated magnetic fields normal to the electron sheet. The
combined effect \cite{peetvasil} of the spatially periodic electrostatic and
magnetic fields of arbitrary shape has been studied \cite
{gerhardts1,gerhardts2}. Analysis of the weak localization and calculation
of the Hall and magneto-resistivities of the 2DEG in a non-homogeneous
magnetic field have been presented \cite{rammer,khaetskii,brey,bykov}.

Recently, different magnetic structures of nanometer scale have been
realized experimentally: a magnetic antidot by depositing a superconducting
disk on top of a 2DEG \cite{geim,smithdot}, a large amplitude magnetic
barriers \cite{jonson,monzon,rpeet1,rpeet2,kubrak1,vancura,kubrak2}\ 
and structures with a
magnetic field alternating in sign \cite{nogaret1} have been produced by a
single or by an array of ferromagnetic lines fabricated on the surface of
the heterostructure in hybrid semiconductor/ferromagnet devices. This
realization of different magnetic regions in an electron gas with sharp
boundaries was a challenge for theoretical studies of the
one-particle electronic states (or the magnetic edge states) moving along
the magnetic interfaces in quantum waveguides \cite{gu}, quantum dots \cite
{sim,reijnpeetmat}, and in infinite 2DEGs \cite{peetrbv,reijnpeet}\ exposed
to a non-homogeneous magnetic field.

The aim of the present paper is to investigate the magnetic edge states and
their transport properties (in the ballistic and diffusive regimes) in a
one-dimensional (1D) channel formed by a parabolic confining potential and
exposed to a normal non-homogeneous magnetic field. Structures are studied
where the magnetic field changes its sign, strength, and both sign and
strength at the magnetic interface. Such a system was recently realized
experimentally \cite{nogaret2} by depositing a ferromagnetic stripe on top
of the electron gas and by applying a background magnetic field normal to
the electron gas.\ Varying the background field results in all the above
situations. We calculate rigorously the energy spectrum and the wave
functions of these systems by matching the general solutions of the
Schr\"{o}dinger equation at the magnetic interface. The corresponding group
velocities along the interface and the particle average position normal to
the interface are obtained. Using the results for the spectrum, we calculate
the conductance and the conductivity in the ballistic and diffusive regimes.

The paper is organized as follows. In Sec. II we present the method we use
to obtain the spectrum. In Sec. III we carry out actual calculations of the
energy spectrum and the wave functions, the group velocity along the
interface and the particle average position normal to the interface for the
three different cases when the magnetic field changes its sign, strength,
and both sign and strength at the magnetic interface. We analyze the
dependence on the confining potential strength and on the magnetic field
strength in one side of the interface while the magnetic field in the other
side is kept fixed. In Sec. IV we calculate the conductance and the
conductivity in the ballistic and diffusive regimes both as a function of
the Fermi energy and of the background magnetic field. The results are
compared with those in case of a homogeneous magnetic field. The results are
summarized in Sec. V.

\section{Approach}

\noindent We investigate the magnetic edge states in a one-dimensional
electron channel along the $y$-direction formed by the parabolic confining
potential $V(x)$ and exposed to a normal non-homogeneous magnetic field $%
B_{z}(x)=B_{1}$ and $B_{z}(x)=-B_{2}$ respectively on the left and the right
hand side of the magnetic interface located at $x=0$ (see Fig.~\ref{fg1}).
This system is placed in a homogeneous background magnetic field $%
B_{z}(x)=B_{b}$. Varying the background magnetic field from $-B_b$ to $B_b$
$(B_b > B_{1},B_{2})$ allows to have situations where the effective
non-homogeneous magnetic field changes its sign, strength, and both sign and
strength at the magnetic interface.

In any finite region along the $x$-direction where the magnetic field is
uniform, the system is described by the single particle Hamiltonian 
\begin{equation}
H=\frac{1}{2m^{\ast }}\left( \overrightarrow{p}+\frac{e}{c}\overrightarrow{A}%
\right) ^{2}+V(x)
\end{equation}
where $m^{\ast }$ is the particle effective mass, $V(x)=m^{\ast }\omega
_{0}^{2}x^{2}/2$ the confining potential with $\omega _{0}$ the confining
potential strength. Because of the system translation invariance in the $y$%
-direction we choose for the vector potential the Landau gauge $%
\overrightarrow{A}=(0,Bx,0)$. In this gauge the Schr\"{o}dinger equation can
be separated with the ansatz 
\begin{equation}
\Psi (x,y)=e^{iky}\psi (x),
\end{equation}
where $\psi $\ is an eigenstate of the one-dimensional problem 
\begin{equation}
\left[ \frac{d}{dx^{2}}+\nu +\frac{1}{2}-\frac{\left( x-X(k)\right) ^{2}}{4}%
\right] \psi \left( x-X(k)\right) =0.  \label{1d}
\end{equation}
Here we introduce the following notations: $\nu +\frac{1}{2}=\left(
\varepsilon -\frac{\hbar ^{2}k^{2}}{2m_{_{B}}}\right) /\hbar \omega ^{\ast }$
is the particle transverse energy in units of the oscillator frequency $%
\omega ^{\ast }=\sqrt{\omega _{B}^{2}+\omega _{0}^{2}},$ $\omega _{B}$ is
the cyclotron frequency, $\varepsilon $ and $k$ are the energy and the
momentum of the particle. The coordinate of the center of orbital rotation
is $X(k)=kl^{\ast }\omega _{{B}}/\omega ^{\ast }$ in units of the length
scale $l{^{\ast }}=\sqrt{\hbar /(m^{\ast }\omega ^{\ast })}$ related to $%
\omega ^{\ast }.$ In the longitudinal direction the electron acquires a new
field dependent mass $m_{_{B}}=m^{\ast }\omega ^{\ast }{}^{2}/\omega _{0}^{2}
$ which is larger than the effective mass $m^{\ast }$. Equation (%
\ref{1d}) is to be solved under the boundary conditions $\psi
(x-X(k))\longrightarrow 0$ when $x\longrightarrow \pm \infty $. The 
solutions are the parabolic cylindrical functions \cite
{abramovic} 
\begin{equation}
D_{\nu }(x)=-\frac{2^{\frac{1}{2}+\frac{\nu }{2}}\,\sqrt{\pi }\,x\,_{1}F_{1}(%
\frac{1}{2}-\frac{\nu }{2},\frac{3}{2},\frac{x^{2}}{2})}{e^{\frac{x^{2}}{4}%
}\,\Gamma (\frac{-\nu }{2})}+\frac{2^{\frac{\nu }{2}}\,\sqrt{\pi }%
\,_{1}F_{1}(\frac{-\nu }{2},\frac{1}{2},\frac{x^{2}}{2})}{e^{\frac{x^{2}}{4}%
}\,\Gamma (\frac{1}{2}-\frac{\nu }{2})},
\end{equation}
with $_{1}F_{1}(a;b;x)$ the Kummer function. For any value of 
$\nu $ there are two independent solutions $D_{\nu }(x)$ and $D_{\nu }(-x)$.
If $\nu \neq 0,1,2,\ldots $, then $D_{\nu }(x)\longrightarrow 0$ when $%
x\longrightarrow +\infty $ and $D_{\nu }(x)\longrightarrow \infty $ when $%
x\longrightarrow -\infty $.

In the non-homogeneous magnetic field case $\nu $, ${X}$ are different on
the left and right hand side of the magnetic interface. We construct the
wave function as 
\begin{equation}
\psi _{\nu _{1},\nu _{2}}(x,{X}_{1},{X}_{2})=\left\{ 
\begin{array}{c}
D_{\nu _{1}}(\sqrt{2}(X_{1}(k)-x)),\text{ if }x<0, \\ 
D_{\nu _{2}}(\sqrt{2}(x-X_{2}(k))),\text{ if }x>0.
\end{array}
\right. 
\end{equation}
Indices $1,2$ refer to the values of quantities for which $\omega ^{\ast }=$ 
$\sqrt{\omega _{B}^{2}+\omega _{0}^{2}}$ is taken with $B=B_{1}$ and $%
B=B_{2},$ respectively. Matching of this wave function and its derivative at 
$x=0$ leads to the following dispersion equation 
\begin{equation}
\left. \frac{d\ln (D_{\nu _{1}}(x-{X}_{1}(k))}{dx}\right| _{x=-0}=\left. 
\frac{d\ln (D_{\nu _{2}}(-x+{X}_{2}(k))}{dx}\right| _{x=+0}.
\label{dispersion}
\end{equation}
By solving this equation we obtain the energy spectrum $\varepsilon _{n}(k)$
and the wave functions $\psi _{\nu _{1},\nu _{2}}(x,{X}_{1},{X}_{2})$ of the
magnetic edge states, which are the solution of the one-dimensional problem
with the effective potential 
\[
V_{eff}(x,k)=\left\{ 
\begin{array}{c}
\frac{1}{2}m\omega _{1}^{\ast }{}^{2}(x-{X}_{1}(k))^{2}+\frac{\hbar ^{2}k^{2}%
}{2m_{_{B_{1}}}},\text{ if }x<0, \\ 
\frac{1}{2}m\omega _{2}^{\ast }{}^{2}(x-{X}_{2}(k))^{2}+\frac{\hbar ^{2}k^{2}%
}{2m_{_{B_{2}}}},\text{ if }x>0.
\end{array}
\right. 
\]
The shape of $V_{eff}(x,k)$ depends strongly on the sign of the wave
number $k$ and on the magnetic field profile.

\section{Spectrum}

\subsection{\noindent Symmetric system: $B_{1}=B_{2},\ sign\left(
B_{1}/B_{2}\right) =-1$}

\noindent In this symmetric case, the dispersion equation (\ref{dispersion})
breaks into two pieces

\begin{equation}
D_{\nu }(x-{X }(k))=0,\ \ \ D_{\nu }^{\prime }(x-{X }(k))=0  \label{zerodwf}
\end{equation}
i.e. the zeroes of the parabolic cylindrical function and its derivative
give the single particle spectrum of the magnetic edge states. Notice that
the first equation gives the spectrum of the usual edge states in a uniform
magnetic field for the infinite hard wall confining potential \cite
{makarov,laughlin,halperin,streda}. First we find the spectrum from the
above Eqs. (\ref{zerodwf}) using the asymptotics of the parabolic cylindric
functions and its derivative in the limits of $k\longrightarrow \pm \infty $.

In the limit of $k\longrightarrow +\infty $ ($x\gg 1,x\gg 2\sqrt[4]{\nu })$
we use the following asymptotic forms for the parabolic cylindric functions 
\cite{gyuninnen}\ and its derivative

\begin{equation}
\left\{ 
\begin{array}{c}
D_{\nu }(-x) \\ 
D_{\nu }^{\prime }(-x)
\end{array}
\right\} \backsim \left\{ 
\begin{array}{c}
1 \\ 
-x
\end{array}
\right\} \left[ x^{\nu }\exp \left( i\pi \nu -\frac{x^{2}}{4}\right) \pm 
\frac{\sqrt{2\pi }}{\Gamma (-\nu )}x^{-\nu -1}\exp \left( \frac{x^{2}}{4}%
\right) \right] ,
\end{equation}
and find the energy

\begin{equation}
\varepsilon _{n}(k)\approx \left( n+\frac{1}{2}\right) \hbar \omega ^{\ast }+%
\frac{\hbar ^{2}k^{2}}{2m_{_{B}}}\mp \frac{2^{n}(kl^{\ast })^{2n+1}}{\sqrt{%
\pi }n!}\left( \frac{\omega _{{B}}}{\omega ^{\ast }}\right) ^{2n+1}\exp
\left( -(kl^{\ast })^{2}\frac{\omega _{{B}}^{2}}{\omega ^{\ast }{}^{2}}%
\right) .  \label{expcorr}
\end{equation}
The corresponding trajectories of the electron orbits have their center of
orbital motion located far from the magnetic interface, i.e. ${X(k)}\gg 1$,
but they
are on the same side of the magnetic interface where the particle is moving.
The energy differs exponentially from the energy of the {\it hybrid} states
Of the uniform magnetic field case. Exponentially small
interaction between the hybrid states located at a finite distance on both
sides of the magnetic interface shifts exponentially small the energy levels up
and down with respect to the bare spectrum of the hybrid states of the
uniform magnetic field case. The wave function of each level of the magnetic
edge states is represented by a curve, which has two peaks. The peaks are
situated far from the magnetic interface both in the positive and the
negative magnetic field regions and are connected with an exponentially
attenuating ''tail'' of the wave function near the interface. In the
two-dimensional case when $\omega _{0}\longrightarrow 0,\
m_{_{B}}\longrightarrow \infty $ and the particle velocity due to the
confining potential becomes zero, the exponential corrections to the
energy results in exponentially small velocities of opposite sign for the
states shifted up and down in energy.\ Notice that the exponential
corrections in Eq. (\ref{expcorr}) cannot be obtained from quasiclassical
considerations.

In the opposite limit of $k\longrightarrow -\infty $ ($\nu \ll 1-x^{2}/2\nu
\ll 1)$ we use the following asymptotic forms for the parabolic cylindric
functions \cite{gyuninnen} and its derivative

\begin{eqnarray}
\left\{ 
\begin{array}{c}
D_{\nu }(x) \\ 
D_{\nu }^{\prime }(x)
\end{array}
\right\}  &\backsim &\left\{ 
\begin{array}{c}
\sqrt{2} \\ 
-x
\end{array}
\right\} \left( \nu +\frac{1}{2}\right) ^{\nu /2}\left( 1-\frac{x^{2}}{2\nu }%
\right) ^{\mp 1/4}\exp \left( -\frac{\nu }{2}-\frac{1}{4}\right)  \\
&\times &\left\{ 
\begin{array}{c}
\cos  \\ 
\sin 
\end{array}
\right\} \left( \frac{2}{3}\nu \left( 1-\frac{x^{2}}{2\nu }\right) ^{3/2}-%
\frac{\pi }{4}\right) ,
\end{eqnarray}
and find the energy 
\begin{equation}
\varepsilon _{n}(k)\approx \left( n+\frac{1}{2}\right) \hbar \omega ^{\ast }+%
\frac{\hbar ^{2}k^{2}}{2m^{\ast }}+\left( a_{n}\frac{kl^{\ast }}{\sqrt{2}}%
\frac{\omega _{B}}{\omega ^{\ast }}\right) ^{2/3},\ \ \ a_{n}=\frac{3\pi }{2}%
\left( n-\frac{1}{2}\mp \frac{1}{4}\right) .
\end{equation}
This spectrum characterizes the particle motion in {\it snake }orbits, i.e.
in trajectories whose center of orbital motion are located far from the
magnetic interface, i.e. ${X(k)}\gg 1,$ but on the opposite side of the 
magnetic interface where the particle is moving. To first order in $k$
the confining potential has no influence on the energy along the $y$%
-direction and the particle mass is the free electron mass $m^{\ast }$. This
is because the effective potential minimum $V_{eff}(x)=\hslash
^{2}k^{2}/2m^{\ast }$ at $x=0$ does not depend on the confining potential
strength $\omega _{0}$ for negative $k$ (see Fig.~\ref{fg2} where we
summarised the three different shapes of the effective 1D potential for the
symmetric system).

The exact spectrum, shown in Fig.~\ref{fg3}, is described by a discrete
quantum number $n=0,1,2,...$ and a continuous momentum $k$. For a given $n$
the energy spectrum exhibits a pronounced asymmetry with respect to positive
and negative values of $k$. The corresponding group velocities $v_{n}$ along
the interface and the particle average semi-thickness $\Delta x_{n}$ normal
to the interface are shown in Fig.~\ref{fg4} (in this symmetrical system, $%
B_{1}=-B_{2}$, the particle average position $\overline{x}_{n}$ is zero, and
therefore we calculated the quantity 
$\Delta x_{n}=\frac{1}{2}\sqrt{\left(x-\overline{x}_{n}\right) ^{2}}$).
It is seen from Figs.~\ref{fg2} and \ref
{fg4} that for large negative values of $k,$ particles are confined in a
narrow region around the magnetic interface. The corresponding wave
functions are represented by one-peak curves localized near the magnetic
interface. These states correspond to the snake orbits, which wiggle around
the magnetic interface moving alternatively in the positive and negative
magnetic field regions. Since the coordinate of the orbit centrum, $X(k),$
increases with $k,$ the radius of the orbit should also increase to ensure
particle motion on the opposite side of the magnetic interface. This
requires an increase of the energy with $k,$ and therefore all these snake
states acquire a large velocity along the interface, which increases
approximately linearly in $k$ while the width $\Delta x_{n}$ of the snake
orbits decreases in $k$ and reaches its minimum value.

For positive values
of $k$ a triangular like barrier is developed at $x=0$ and the effective
potential becomes the double well (see Fig.~\ref{fg2}). The height of the
barrier is $V_{eff}(x)=\hslash ^{2}k^{2}/2m^{\ast }$ at $x=0$ while the well
minima are $\hslash ^{2}k^{2}/2m_{_{B}}$ at $x=\pm X(k)$. For small positive
values of $k$ the particle motion is still snake-like around the magnetic
interface. The momentum and the velocity of these states have opposite sign.
Starting from some value $k_{n}>0$ the ground state electron wave functions
is split into the two peaks by the barrier. For large positive values
of $k$ the spectrum characterizes the exponentially weak coupled two hybrid
states located at $x\approx \pm X(k)$ in the positive and negative magnetic
field regions and therefore rotate in opposite direction. The velocity of
these hybrid states is mainly due to the confining potential and directed
opposite to the velocity of the snake states. The absolute value of the
velocity is determined by the height of the minima of the effective
potential wells, i.e. by the mass $m_{_{B}}$ which is larger than the free
electron mass $m^{\ast }$, therefore the snake states are faster than the
hybrid states. For large positive values of $k$ both the group velocities $%
v_{n}$ and the particle average semi-thickness $\Delta x_{n}$ increase
approximately linearly in $k$ (see Fig.~\ref{fg4}). The velocity (the
average semi-thickness) of the symmetric and anti-symmetric states tend
respectively to its asymptotic value $v_{n}=\hslash k/m_{_{B}}$ ($\Delta
x_{n}=X$) from above (below) and below (above).

\subsection{\noindent Asymmetric system: $B_{1}\neq B_{2},\ sign\left(
B_{1}/B_{2}\right) =-1$}

\noindent The effective potential for this asymmetric system where the
magnetic field changes both its strength and sign at the magnetic interface
is shown in Fig.~\ref{fg5}. In this case the effective potential $%
V_{eff}(x,k)$ exhibits a pronounced asymmetry both as a function of $k$ and $%
x$. For negative values of $k,$ the effective potential is a triangular-like
asymmetric well with a minimum of $V_{eff}(x)=\hslash ^{2}k^{2}/2m^{\ast }$
at $x=0.$ For positive values of $k$ the effective potential is a double
well with different minima $\hslash ^{2}k^{2}/2m_{_{B_{1}}}$and $\hslash
^{2}k^{2}/2m_{_{B_{2}}}$ at the positions $x=+X_{1}(k)$ and $x=-X_{2}(k)$,
respectively. The triangular like barrier between the wells has again the
height $V_{eff}(x)=\hslash ^{2}k^{2}/2m^{\ast }$ at $x=0.$ Thus the
confining potential together with the non-homogeneous magnetic field induces
three effective masses ($m^{\ast }$ for negative and $m_{_{B_{1}}}$, $%
m_{_{B_{2}}}$ for positive values of $k$) in the system. For negative values
of $k$, the spectrum corresponds to snake orbits with free-like motion
and with mass $m^{\ast }$ along the $y$-direction (see Fig.~ \ref{fg6}).\
These states are effectively localized in the vicinity of the magnetic
interface in the region where the magnetic field is smaller and the magnetic
length is larger. The group velocity is approximately linear (see Fig.~\ref
{fg7}) and the particle average position $\overline{x}_{n}$ is approximately
independent of the wave number (see Fig.~\ref{fg8}). The $n=1$ level is the
closest to the magnetic interface and most remote from the $n>1$ states. For
positive $k$ the spectrum characterizes the hybrid states with two different
masses $m_{_{B_{1}}}$ and $m_{_{B_{2}}}$. Each energy band $n$ has $n$
anti-crossings with these hybrid states. For some positive value of $k$ the
group velocity $v_{n}$ and the particle average position $\overline{x}_{n}$
start to oscillate as a function of the wave number and the particle tunnels
periodically from the left to the right side of the quantum wire and vice
versa (see Fig.~\ref{fg8}).
At $k\rightarrow +\infty $ all states tend to be localized in the
region where the magnetic field is large and the well of the effective
potential is lower. Contrary to the symmetric system, the ground state wave
function consists now of a curve with one peak. At the
anti-crossing points the wave function changes sign and its peak position
shifts rapidly by changing its sign and value (see Fig.~\ref{fg9}). This
corresponds to a tunneling of the particle from the one well of the
effective potential to the other. This picture is true even if there is only
a small difference between the magnetic field on both sides of the
interface, which brakes the symmetry of the system and the particle is
forced to choose one of the wells.

\subsection{\noindent Asymmetric system: $B_{1}\neq B_{2},\ sign\left(
B_{1}/B_{2}\right) =1$}

\noindent In such systems in which at the magnetic interface the magnetic
field changes only its strength, the effective potential consists of only 
one well (see Fig.~\ref{fg10}). For negative values of $k$ the well is 
located at $x=-X_{2}(k)$ with minimum value $\hslash ^{2}k^{2}/2m_{_{B_{2}}}$.
It is higher and broader than the well for positive values of $k$ with 
minimum $\hslash^{2}k^{2}/2m_{_{B_{1}}}$ at the position $x=+X_{1}(k)$
($B_{1}>B_{2}$). There
are no snake states in this system and the spectrum for both large negative
and positive values of $k$ characterizes hybrid states with masses $%
m_{_{B_{2}}}$ and $m_{_{B_{1}}}$ (Fig.~\ref{fg11}). These states rotate in
the same direction on both sides of the magnetic interface and their group
velocities have opposite sign. Both the velocity and the particle average
position of any level $n$ are approximately linear in $k$ for large negative
and positive $k$ and tend respectively to their asymptotic values $%
v_{n}=\hslash k/m_{_{B_{2}}}$, $\overline{x}_{n}=-{X }_{2}(k)$ and $%
v_{n}=-\hslash k/m_{_{B_{1}}},$ $\overline{x}_{n}=X_{1}(k)$, which are
independent of $n$ (see Fig.~\ref{fg12}). For intermediate values of $k$ the
velocity and the particle average position exhibit smooth oscillations as a
function of $k.$ For these values of $k$ the corresponding states are
confined in the effective potential which consists of two partial parabolas
with different strengths. With varying $k$ the influence of each parabola
changes strongly in contributing to these states.

\subsection{\noindent Dependence on the confining potential strength}

\noindent In Figs.~\ref{fg13}\ (a-d) the dependence of the energy on the
confining potential frequency $\omega _{0}$ is depicted when $B_{1}=-3B_{2}$
and $B_{1}=3B_{2}$ both for $kl^{\ast }=\pm 2.$ It is seen that there is a
strong asymmetry with respect to the sign of the magnetic field and of the
wave number. For the systems where the magnetic field changes its sign and
strength at the magnetic interface$,$ the states with $kl^{\ast }=-2$
correspond to snake orbits. In this case the dependence on $\omega _{0}$ is
very weak because the effective potential consists of only one well with
minimum value $V(0)=\hbar ^{2}k^{2}/2m^{\ast }$, as we mentioned above,
which does not depend on $\omega _{0}.$ Varying $\omega _{0}$ changes only
the sharpness of the banks of the potential well, which results in a much
weaker dependence of the energy bands on $k$. In the case of $kl^{\ast }=+2$
(this corresponds to the region of two hybrid states with different masses $%
m_{_{B_{1}}}$ and $m_{_{B_{2}}}$) the energy strongly depends on $\omega
_{0}.$ For $\omega _{0}=0$ we have a two-dimensional system \cite
{peetrbv,reijnpeet} and some states are very close in energy due to the
special choice of the ratio of the two magnetic fields $B_{1}/B_{2}$ which 
is an
integer$.$ For small values of $\omega _{0},$ the energy increases strongly
with $\omega _{0}.$ Several anti-crossings appear between the energy bands
with different $n$ for this choice of parameters. With further increase of $%
\omega _{0}$ the increase of energy becomes weaker and the simple
oscillatory states with $B_{1,2}\approx 0$ correspond to the limit $\omega
_{0}/\omega _{B}\gg 1.$

In the systems where the magnetic field changes only its strength at the
magnetic interface, both $kl^{\ast }=+2$ and $kl^{\ast }=-2$ correspond to
hybrid states on the left and right side of the interface, respectively, and
the energy dependence on $\omega _{0}$ is qualitatively similar for these 
states. The quantitative difference is a result of the different values of 
the masses 
$m_{_{B_{1}}}$ and $m_{_{B_{2}}}$ (e.g. $m_{_{B_{1}}}/m_{_{B_{2}}}=45/13$
for $\omega_{B_{1}}=2\omega_0=3\omega_{B_{2}}$). 
It is easy to see that for $\omega _{0}=0$ the energy takes values
near $1/6,1/2,5/6,...$ if $kl^{\ast }=-2$ and $1/2,3/2,5/2,...$ $\ $if $%
kl^{\ast }=+2$ as it should be for a two-dimensional system 
\cite{peetrbv,reijnpeet}.

\subsection{\noindent Dependence on the magnetic field $B_{2}$}

\noindent In Figs.~\ref{fg14}\ (a,b) we plot the energy dependence on the
magnetic field $B_{2}$ for fixed $B_{1}$ and for the confining potential
frequency $\omega _{0}/\omega _{{B}_{1}}=1/2$ for $kl^{\ast }=\pm 2.25.$
Notice there is a strong asymmetry with respect to the sign of $k$. When $%
kl^{\ast }=-2.25$ the energy dependence on $B_{2}$ for negative $B_{2}$ is
weaker than for positive $B_{2}$ because in the first range the spectrum
characterizes the snake states and the dependence on $B_{2}$ is mainly due
to the dependence of $\omega _{2}^{\ast }$ on $B_{2}$ while in the second
range the spectrum characterizes the hybrid states and the energy dependence
is due to the dependence of both $\omega _{2}^{\ast }$ and $m_{_{B_{2}}}$ on 
$B_{2}.$ It is easy to see that for $B_{2}$/$B_{1}=\pm 1,\pm 1/3$ the energy
values are consistent with that in Figs.~\ref{fg3},\ref{fg6}, and \ref{fg11}.

For $kl^{\ast }=+2.25$ the first energy level almost does not depend on $B_{2}$.
For positive values of $B_{2}$ the spectrum describes only the hybrid states
and for the chosen large value of $kl^{\ast }=+2.25$ the energy equals 
approximately its asymptotic value, 
$\hbar \omega _{1}^{\ast }/2+\hbar ^{2}k^{2}/2m_{_{{B}_{1}}}$, which is 
independent of $B_{2}$. For negative values of $B_{2},$ the energy is again 
approximately those of the hybrid states with the same energy because
$kl^{\ast}=+2.25$ is larger than the value $k_{1}l^{\ast }$ at which the ground 
state exhibits an anti-crossing. Analogous behavior is found for the second 
energy level, but now starting from some negative value of $B_{2}$ this level,
after anti-crossing with the level $n=3,$ tends to the first level because
of the degeneracy of the symmetric and anti-symmetric terms in the $%
B_{1}=-B_{2}$ symmetric system (see Fig.~\ref{fg3})$.$ For $B_{2}<-B_{1},$
the degeneracy is lifted and the energy increases with $|B_{2}|.$

\section{Transport}

\noindent We calculate the zero temperature two terminal magneto-conductance 
for a perfect
conductor using the B\"{u}ttiker formula \cite{but} 
\begin{equation}
G(E_{F})=\frac{2e^{2}}{\hbar }N(E_{F}),
\end{equation}
where $N(E)$ is the number of magnetic edge states with energy $E$ and 
positive velocity.
From Fig.~\ref{fg15} it is seen that the conductance, in the ballistic regime
And for different magnetic field profiles, exhibits stepwise variations as a
function of the Fermi energy. For a given energy and confining potential
strength, the conductance in the non-homogeneous magnetic field is nearly
twice that for the homogeneous field case. The conductance decreases when going from 
the profile $B_{1}=-3B_{2}$ to the profiles $B_{1}=-B_{2}$, and $B_{1}=+3B_{2}$.
For the symmetric profile the narrow plateau
is followed by broad ones. This asymmetry is not visible for the other
profiles. The conductance is the same in the positive and negative
directions along the magnetic interface despite the strong asymmetry in the
magnitude of the velocity of the states moving in opposite directions.

The conductivity in the diffusive regime is calculated in the relaxation
time approximation 
\begin{equation}
\sigma _{_{1D}}=\frac{e^{2}}{L}\sum\limits_{n,k}v_{n}^{2}(k)\tau \left(
\varepsilon \right) \left( -\frac{\partial f_{T}}{\partial \varepsilon }%
\right) ,  \label{conductivity}
\end{equation}
where $L$ is the length of the quantum wire, $\tau $ is the momentum
relaxation time, $f_{T}$ is the Fermi-Dirac distribution function at
temperature $T.$ We calculate the conductivity in the zero temperature
limit. Then the derivative of the Fermi function is a $\delta $-function.
Replacing all quantities in Eq. (\ref{conductivity}) by their values at the
Fermi energy, we obtain 
\begin{equation}
\left. \sigma _{_{1D}}=\frac{2e^{2}}{h}\tau
(E_{F})\sum\limits_{n}|v_{n}(k)|\right| _{\varepsilon =E_{F}}.
\end{equation}
In the diffusive regime, we have calculated separately the conductivity due
to states with negative and positive velocities as a function of the Fermi
energy for the two magnetic field profiles, $B_{1}=-3B_{2}$ and $%
B_{1}=+3B_{2}$ (see Fig.~\ref{fg16}). In both cases the conductivity due to
states with negative velocities (dashed curves) is larger than that due to
states with positive velocities (dotted curves). In the case when the
magnetic field changes its sign, the states with negative velocities are the
snake states, which are always faster than the states with positive
velocities which are related to the hybrid states. In the case of $%
B_{1}=+3B_{2}$ all the states are hybrid states, however, the contribution
to the conductivity of the states with negative velocities is larger because
these states are located in a region with small magnetic field, have
the small mass $m_{_B},$ and large velocity $v_{n}$. For both $v_{n}>0$
and $v_{n}<0$ parts, the conductivity has an oscillating structure as a 
function of the Fermi energy which is due to a divergence of the density of 
states at the bottom of the $\varepsilon_n(k)$ band. However, the 
contributions due to states with $v_{n}>0$ exhibit
an additional structure related to the oscillations of the group velocity as
a function of $k$. This structure is more pronounced in the case of $%
B_{1}=-3B_{2}$, the conductivity has additional distinct minima that reflect
the tunneling effect discussed above. Notice that the conductivity of the
system with the magnetic field profile $B_{1}=-3B_{2}$ is roughly $1.5$ times 
larger than that for the field profile $B_{1}=+3B_{2}.$

In Fig.~\ref{fg17} the magnetic depopulation diagram is plotted as a
function of the background magnetic field $B_{b}$ and the wave number $k$
for the initial magnetic field profile $B_{1}=2B_{0}=-3B_{2}$ ($B_{0}$ is
the resonance field for which $\omega _{0}=\omega _{{B}_{0}}$) and for the
Fermi energy $E_{F}=5\hbar \omega _{1}^{\ast }$. In the shaded region the
effective magnetic field changes its sign at the magnetic interface. At the
background magnetic field $B_{b}=0$ there are $18$ current carrying states
represented by the solid dots in the figure$.$ The left $9$ symbols
correspond to snake states while the right $9$, to hybrid states
with both $m_{_{B_{1}}}$ and $m_{_{B_{2}}}$ masses. 
The maximum number of current carrying states, $20$, is achieved in the
small region of the background magnetic field around $B_{sym}=-2/3B_{0}$
where the effective magnetic field on both sides of the magnetic interface
has equal strength, $4/3B_{0},$ and opposite sign. The background magnetic
field dependence is symmetric with respect to this point. Out of the shaded
region the current carrying states are only the hybrid states. At the edges
of the shaded region, the effective magnetic field becomes zero on either
the left or the right side of the magnetic interface. When the absolute
value of the background magnetic field increases starting from the value $%
B_{sym}$, the number of current carrying states decreases monotonically.

In Figs.~\ref{fg18} and \ref{fg19} we plot the magneto-resistance as a
function of the background magnetic field in the ballistic and diffusive
regimes, respectively, for the situation corresponding to Fig.~\ref{fg17}.
The resistance in the ballistic regime exhibits stepwise variations as a
function of $B_{b}$ and has a minimum at $B_{b}=B_{sym},$ which is shifted
with respect to the minimum of the resistance in the homogeneous magnetic
fields (thin dashed curve). In the later case the dependence on $B_{b}$ is
stronger because the number of current carrying states is smaller. In the
diffusive regime the resistance exhibits small peaks as a function of $B_{b}$%
\ that are associated with the magnetic depopulation effect and that are on
top of a positive magneto-resistance background, which increases with $B_{b}$
when $B_{b}$ has the same sign as the initial magnetic field of the region
where the magnetic field is larger. For small values of $B_{b}$ the
resistance in the homogeneous field is smaller. The slope of the resistance
variation in $B_{b}$ in the homogeneous field case is larger than for 
non-homogeneous
fields. Notice that the minima of the conductivity due to the hybrid states
with $v_{n}>0$ associated with the tunneling effect (see Fig.~\ref{fg16})
are not visible in the background magnetic field dependence of the
resistance in Fig.~\ref{fg19}. This is possibly due to the special choice of
the initial parameters ($B_{1}=2B_{0}=-3B_{2})$ and of the Fermi energy, the
possible peak values of the resistance are out of the small range of the
variation of $B_{b}$ (from $0$ to $2/3B_{0})$ where the effective magnetic
field changes its sign. Moreover, the effective magnetic field on one side
of the magnetic interface is given by $-B_{2}+B_{b},$ i.e. the increase of
the background magnetic field diminishes the effective magnetic field,
namely in the region where the initial magnetic field is smaller

\section{Summary}

We developed a theory for the non-homogeneous magnetic field induced
magnetic edge states and their transport in a quantum wire formed by a
parabolic confining potential. We studied systems in which the magnetic
field perpendicular to the wire axis exhibits a discontinuous jump in the
transverse direction and changes its sign, strength, and both sign and
strength at the magnetic interface. The energy spectrum and the wave
functions of the magnetic edge states were calculated by matching the
general solutions of the Schr\"{o}dinger equation at the magnetic interface.
The corresponding group velocities along the interface and the particle
average position normal to the interface were obtained.

The spectrum consists of alternating symmetrical and anti-symmetrical terms
and is described by a discrete quantum number $n=0,1,2,...$ and the 
momentum $k$ along the wire.
For given $n$ the energy spectrum exhibits a pronounced asymmetry with
respect to positive and negative values of $k$ and describes snake orbits
and hybrid states. Contrary to two-dimensional systems \cite
{peetrbv,reijnpeet}, the confining potential together with the
non-homogeneous magnetic field induces three effective masses, which can
account for most of the system properties. When the magnetic field changes
its sign and strength, all states with negative momenta (the snake orbits)
are effectively localized in the vicinity of the magnetic interface in the
region where the magnetic field is small. The group velocity is
approximately linear and the particle average position is approximately
independent of the momentum. For a positive momentum the spectrum exhibits
anti-crossings, the group velocity and the particle average position
oscillate as a function of the momentum and the particle tunnels
periodically from the left to the right side of the quantum wire and vice
versa. At $k\rightarrow +\infty $ all states tend to be localized in the
region where the magnetic field is large.

The conductance in the ballistic regime exhibits stepwise variations as a
function of the Fermi energy and of the background magnetic field$.$ For a
given energy and confining potential strength, the conductance in the
non-homogeneous magnetic field is nearly twice that in the case
of a homogeneous field. The conductance has a maximum as a function of $%
B_{b} $ at the value for which the effective magnetic field on the left and
on the right hand side of the magnetic interface has the same strength and
opposite sign.

In the diffusive regime, we calculated separately the conductivity for
negative and positive velocities as a function of the Fermi energy and the
background magnetic field $B_{b}$. The conductivity due to states with
negative velocities is large. The conductivity oscillates as a
function of the Fermi energy. The contributions due to states with positive
velocities exhibit an additional structure related to the oscillations of
the group velocity as a function of $k$. In the systems where the magnetic
field changes its sign this structure is more pronounced with additional
distinct minima related to the tunneling of the particle between different
magnetic regions. The resistance exhibits small peaks as a function of
background magnetic field\ that are associated with magnetic
depopulation effects.

\section*{Acknowledgement}

This work was partially supported by the Flemish Science Foundation
(FWO-Vl), the Inter-university Micro-Electronic Center (IMEC, Leuven), 
the "Onderzoeksraad van de Universiteit Antwerpen", and the IUAP-IV (Belgium).
S. M. B. was supported by a DWTC-fellowship to promote the S \& T collaboration
with Central and Eastern Europe and a CRDF grant No 375100. 
We acknowledge fruitful discussions with J. Reijniers.

\begin{figure}[tbp]
\caption{Schematic diagram of the system under study. The quantum wire is
formed along the $y$-axis by a parabolic confining potential $V(x)=m^*\protect%
\omega_0^2x^2/2$ in the $x$-direction. In the $z$-direction a
non-homogeneous magnetic field $B_z=B_1$ and $B_z=-B_2$ is applied
respectively on the left and the right hand side of the magnetic interface
at $x=0$. A homogeneous background magnetic field, $B_z=B_b$, can be
additionally applied.
The channel width is $W$, the length scale $l_0^2=\hbar/(m^*\protect\omega_0)$
 is related to the confining potential strength $\protect\omega_0$,
 $m^*$ is the electron
effective mass, $E_F$ and $k_F$ are the Fermi energy and momentum,
respectively.}
\label{fg1}
\end{figure}

\begin{figure}[tbp]
\caption{The effective potential $V_{eff}(x,k)$ for the symmetric $B_1=-B_2$
magnetic field profile. The classical trajectories are shown schematically
for the snake orbits ($k=0,-1.5$) and for the hybrid states ($k=1.5$).}
\label{fg2}
\end{figure}

\begin{figure}[tbp]
\caption{The energy spectrum for the $8$ lowest bands corresponding to the
situation of Fig.~2.}
\label{fg3}
\end{figure}

\begin{figure}[tbp]
\caption{The group velocity along the magnetic interface (the thin curves,
the right and top axes) and the particle average semi-thickness (the thick
curves, the left and bottom axes) corresponding to Fig.~3.}
\label{fg4}
\end{figure}

\begin{figure}[tbp]
\caption{The effective potential $V_{eff}(x,k)$ for the asymmetric 
$B_1=-3B_2$ magnetic field profile. $V_{eff}(x,k)$ is a single well for $k<0$
and a double well for $k>0$. All wells have different widths and are shifted
up for different heights determined by the masses $m^*$, $ m_{_{B_{1}}}$, and 
$ m_{_{B_{2}}}$. The classical trajectories are shown schematically for the
snake orbits ($k=0,-1.5$) and for the hybrid states ($k=1.5$).}
\label{fg5}
\end{figure}

\begin{figure}[tbp]
\caption{The energy spectrum for the $8$ lowest bands corresponding to the
situation of Fig.~5. Each energy band $n$ has $n$ anti-crossings. The
symbols (solid dots) correspond to the classical trajectories for $k=1.5$ in
Fig.~5.}
\label{fg6}
\end{figure}

\begin{figure}[tbp]
\caption{The group velocity along the magnetic interface corresponding to
Fig.~6. There are three preferential directions in this figure, $\protect%
\theta$ and $\protect\theta_1$, $\protect\theta_2$, which define the
asymptotic velocities of the snake orbits and hybrid states with the masses $%
m^*$ and $ m_{_{B_{1}}}$, $ m_{_{B_{2}}}$, respectively. The symbols correspond to
the classical trajectories for $k=1.5$ in Fig.~5.}
\label{fg7}
\end{figure}

\begin{figure}[tbp]
\caption{The particle average position normal to the magnetic interface
corresponding to the $5$ lowest energy bands in Fig.~6. The symbols
correspond to the classical trajectories for $k=1.5$ in Fig.~5.}
\label{fg8}
\end{figure}

\begin{figure}[tbp]
\caption{The particle wave functions corresponding to the $2$ lowest energy
bands in Fig.~6 for several values of the wave number $k$ for $n=1$ (a) and $%
n=2$ (b).}
\label{fg9}
\end{figure}

\begin{figure}[tbp]
\caption{The effective potential $V_{eff}(x,k)$ for the asymmetric magnetic
field profile $B_1=+3B_2$. $V_{eff}(x,k)$ is always a single well with
different heights and widths for $k<0$ and $k>0$. The classical trajectories
for the hybrid states at $k=0,\pm1.5$ are shown schematically.}
\label{fg10}
\end{figure}

\begin{figure}[tbp]
\caption{The energy spectrum for the $7$ lowest bands corresponding to the
situation of Fig.~10.}
\label{fg11}
\end{figure}

\begin{figure}[tbp]
\caption{The group velocity along the magnetic interface (the thick curves,
the right and top axis) and the particle average position (the thin curves,
the left and bottom axis) corresponding respectively to the $7$ and $5$
lowest energy bands in Fig.~11.}
\label{fg12}
\end{figure}

\begin{figure}[tbp]
\caption{The dependence of the energy on the confining potential strength
for different magnetic field profiles and for different values of the
wave number.}
\label{fg13}
\end{figure}

\begin{figure}[tbp]
\caption{The dependence of the energy on the magnetic field $B_2$ for fixed $%
B_1$ and for different values of the wave number.}
\label{fg14}
\end{figure}

\begin{figure}[tbp]
\caption{Dependence of the conductance on the Fermi energy in the ballistic
regime for different magnetic field profiles and for fixed confining
potential strength.}
\label{fg15}
\end{figure}

\begin{figure}[tbp]
\caption{Dependence of the conductivity, in units of
$\protect\sigma_0=e^2\protect\tau/(\protect\pi m^{\ast} l^{\ast})$,
on the Fermi energy in the diffusive
regime for different magnetic field profiles and for fixed confining
potential strength. The contributions to the conductivity of the states with
positive and negative velocity are shown separately.}
\label{fg16}
\end{figure}

\begin{figure}[tbp]
\caption{The magnetic depopulation diagram as a function of the background
magnetic field and the wave number for the initial magnetic field profile $%
B_{1}=-3B_{2}$ and for a fixed Fermi energy $E_F=5\sqrt{5}\hbar\omega_0$,
$l^2_{0}=\hbar /(m^{\ast }\protect\omega _{0})$. The solid dots correspond to
the current carrying states at $B_{b}=0$. The shaded region corresponds to 
systems where the effective magnetic field changes its sign at the magnetic 
interface.}
\label{fg17}
\end{figure}

\begin{figure}[tbp]
\caption{Dependence of the resistance on the background magnetic field 
in the ballistic regime for the $B_1=-3B_2$ initial magnetic field profile 
and a
fixed Fermi energy $E_F=5\sqrt{5}\hbar\omega_0$. The resistance in case of
a homogeneous magnetic field is plotted for comparison.}
\label{fg18}
\end{figure}

\begin{figure}[tbp]
\caption{Dependence of the resistance, in units of 
$R_{0}=\protect\pi m^{\ast }l_{0}/(e^{2}\protect\tau )$ with
$l^2_0=\hbar/(m^*\omega_0)$, 
on the background magnetic field in the diffusive regime for the 
$B_{1}=-3B_{2}$ initial magnetic field profile and
a fixed Fermi energy $E_F=5\sqrt{5}\hbar\omega_0$. The resistance in the 
homogeneous magnetic field case is plotted for comparison.}
\label{fg19}
\end{figure}


\begin{references}
\bibitem[\ast]{smb} Permanent address: Department of Radiophysics, Yerevan State
University, 375049 Yerevan, Armenia

\bibitem[o]{fmp} Electronic mail: peeters\@uia.ua.ac.be

\bibitem{landwehr92}  {\em High Magnetic Fields in Semiconductor Physics III}%
, edited by G. Landwehr (Springer-Verlag, Berlin Heidelberg, 1992).

\bibitem{peetershb}  F.~M. Peeters and J.~De Boeck, in {\em Handbook of
nanostructured materials and technology}, edited by N.~S. Nalwa, Vol.~3
(Academic Press, N. Y., 1999), p. 345.

\bibitem{ye1}  P.~D. Ye, D. Weiss, R. R. Gerhardts, M. Seeger, K. von Klitzing, 
K. Eberl, and H. Nickel, Phys. Rev. Lett. {\bf 74}, 3013 (1995).

\bibitem{izawa}  S. Izawa, S. Katsumoto, A. Endo, and Y. Iye, J. Phys. Soc.
Jpn. {\bf 64}, 706 (1995).
\bibitem{carmona} H. A. Carmona, A. K. Geim, A. Nogaret, P. C. Main, 
T. J. Foster, M. Henini, S. P. Beaumont, and M. E. Blamire, Phys. Rev. Lett.
{\bf 74}, 3009 (1995).
\bibitem{foden1}  C.~L. Foden, M.~L. Leadbeater, J.~H. Burroughes, and M.
Pepper, J. Phys.: Cond. Matt. {\bf 6}, L127 (1994).

\bibitem{dubrovin}  B.~A. Dubrovin and S.~P. Novikov, Sov. Phys. JETP {\bf 52%
}, 511 (1980).

\bibitem{vilms}  P.~P. Vil'ms and M.~V. \'{E}ntin, Sov. Phys. Semicond. {\bf %
22}, 1209 (1988).

\bibitem{peetvasil}  F.~M. Peeters and P. Vasilopoulos, Phys. Rev. B 
{\bf 47}, 1466 (1993).

\bibitem{yoshnag}  H. Yoshioka, J. Phys. Soc. Jpn. {\bf 59}, 2884 (1990).

\bibitem{avishai}  Y. Avishai and Y.~B. Band, Phys. Rev. B {\bf 40},
3429 (1989).

\bibitem{peetmat}  F.~M. Peeters and A. Matulis, Phys. Rev. B {\bf 48}%
, 15 166 (1993).

\bibitem{muller}  J.~E. M\"{u}ller, Phys. Rev. Lett. {\bf 68}, 385 (1992).

\bibitem{hofstetter}  E. Hofstetter, J.~M.~C. Taylor, and A. MacKinnon,
Phys. Rev. B {\bf 53}, 4676 (1996).

\bibitem{takagaki}  Y. Takagaki and K. Ploog, Phys. Rev. B {\bf 53},
3885 (1996).

\bibitem{calvo}  M. Calvo, Phys. Rev. B {\bf 48}, 2365 (1993).

\bibitem{peetmatvas}  A. Matulis, F.~M. Peeters, and P. Vasilopoulos, Phys.
Rev. Lett. {\bf 72}, 1518 (1994).

\bibitem{ibrahim2}  I.~S. Ibrahim and F.~M. Peeters, Phys. Rev. B {\bf %
52}, 17 321 (1995).

\bibitem{gerhardts1}  R.~R. Gerhardts, Phys. Rev. B {\bf 53}, 11064
(1996).

\bibitem{gerhardts2}  S.~D.~M. Zwerschke, A. Manolescu, and R.~R. Gerhardts,
Phys. Rev. B {\bf \ 60}, 5536 (1999).

\bibitem{rammer}  J. Rammer and A.~L. Shelnakov, Phys. Rev. B {\bf 36}%
, 3135 (1987).

\bibitem{khaetskii}  A.~V. Khaetskii, J. Physics Cond. Matt. {\bf 3}, 5115
(1991).

\bibitem{brey}  L. Brey and H.~A. Fertig, Phys. Rev. B {\bf 47}, 15 961 (1993).

\bibitem{bykov}  A.~A. Bykov, G.~M. Gusev, J.~R. Leite, A.~K. Bakarov, N.~T.
Moshegov, M. Casse, D.~K. Maude, and J.~C. Portal, Phys. Rev. B {\bf 61%
}, 5505 (2000).

\bibitem{geim}  A.~K. Geim, I.~V. Grigorieva, S.~V. Dubonos, J.~G.~S. Lok,
J.~C. Maan, A.~E. Filipov, and F.~M. Peeters, Nature {\bf 390}, 259 (1997).

\bibitem{smithdot}  A.~Smith R. Taboryski, L.~T. Hansen, C.~B. Sorensen, P.
Hedegard, and P.~E. Lindelof, Phys. Rev. B {\bf 50}, 14726 (1994).

\bibitem{jonson}  M. Jonson, B.~R. Bennett, M.~J. Yang, M.~M. Miller, and
B.~V. Shanabrook, Appl. Phys. Lett. {\bf 71}, 974 (1997).


\bibitem{monzon} F.~G. Monzon, M. Jonson, and M.~L. Roukes, Appl. Phys. Lett. {\bf 71}, 3087 (1997).

\bibitem{rpeet1}  J. Reijniers and F.~M. Peeters, Appl. Phys. Lett. {\bf 73},
357 (1998).

\bibitem{rpeet2}  J. Reijniers and F.~M. Peeters, J. Appl. Phys. {\bf 87},
8088 (2000).

\bibitem{kubrak1}  V. Kubrak, F. Rahman, B.~L. Gallagher, P.~C. Main,
M. Henini, C.~H. Marrows, and M.~A. Howson, Appl. Phys. Lett. {\bf 74}, 2507
(1999). 

\bibitem{vancura}  T. Van\v{c}ura, T. Ihn, S. Broderick, K. Ensslin, W.
Wegscheider, and M Bichler, Phys. Rev. B {\bf 62}, 5074 (2000).

\bibitem{kubrak2}  V. Kubrak A.~C. Neumann, B.~L. Gallagher, P.~C. Main,
M. Henini, C.~H. Marrows, and M.~A. Howson, Physica E {\bf 6}, 755 (2000).

\bibitem{nogaret1}  A. Nogaret S. Carlton, B.~L. Gallagher, P.~C. Main,
M. Henini, R. Writz, R. Newbury, M.~A. Howson, and S.~P. Beaumont, Phys.
Rev. B {\bf 55}, R16037 (1997). 

\bibitem{gu}  B.~Y. Gu, W.~D. Sheng, X.~H. Wang, and J. Wang, Phys.
Rev. B {\bf 56}, R13434 (1997).

\bibitem{sim}  H.~S. Sim, K.~H. Ahn, K.~J. Chang, G. Ihm, N. Kim, and
S.~J. Lee, Phys. Rev. Lett. {\bf 80}, 1501 (1998).

\bibitem{reijnpeetmat}  J. Reijniers, F.~M. Peeters, and A. Matulis,
Phys. Rev. B {\bf 59}, 2817 (1999).

\bibitem{peetrbv}  F.~M. Peeters, J. Reijniers, S.~M. Badalian, and P.
Vasilopoulos, Microelectronic Engineering {\bf 74}, 405 (1999).

\bibitem{reijnpeet}  J. Reijniers and F.~M. Peeters, J. Phys.: Cond. Matt.
(2000).

\bibitem{nogaret2}  S.~J. A.~Nogaret, Bending, and M. Henini, Phys.
Rev. Lett. {\bf 84}, 2231 (2000).

\bibitem{abramovic}  M. Abramowitz and I.~A. Stegun, {\em Handbook of
Mathematical Functions} (Nauka, Moscow, 1979), p. 494.

\bibitem{makarov}  E.~A. Kaner, N.~M. Makarov, and I.~M. Fuks, Sov.
Phys. JETP {\bf 28}, 483 (1969).

\bibitem{laughlin}  R.~B. Laughlin, Phys. Rev. B {\bf 23}, 5632 (1981).

\bibitem{halperin}  B.~I. Halperin, Phys. Rev. B {\bf 25}, 2185 (1982).

\bibitem{streda}  A.~H. MacDonald and P. St\v{r}eda, Phys. Rev. B {\bf 29}, 
1616 (1984).

\bibitem{gyuninnen}  E.~M. Gyuninnen and G.~I. Makarov, in {\em %
Problems of Diffraction and Propagation of Waves}, edited by E.~M. Gyuninnen
(Leningrad State University Press, Leningrad, 1962), Vol.~1, p. 24.

\bibitem{but}  M. B\"{u}ttiker, Phys. Rev. B {\bf 38}, 9375 (1988).
\end{references}
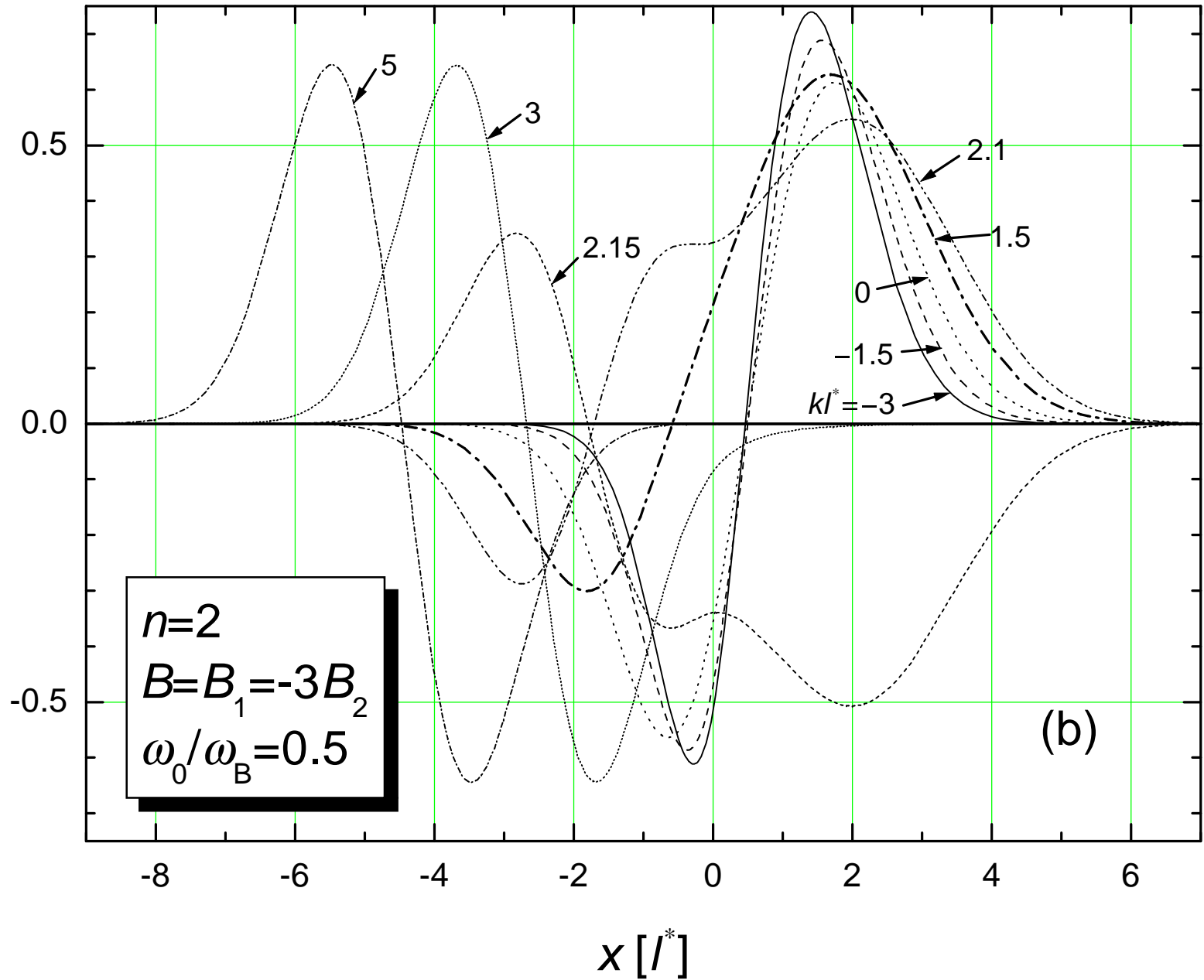
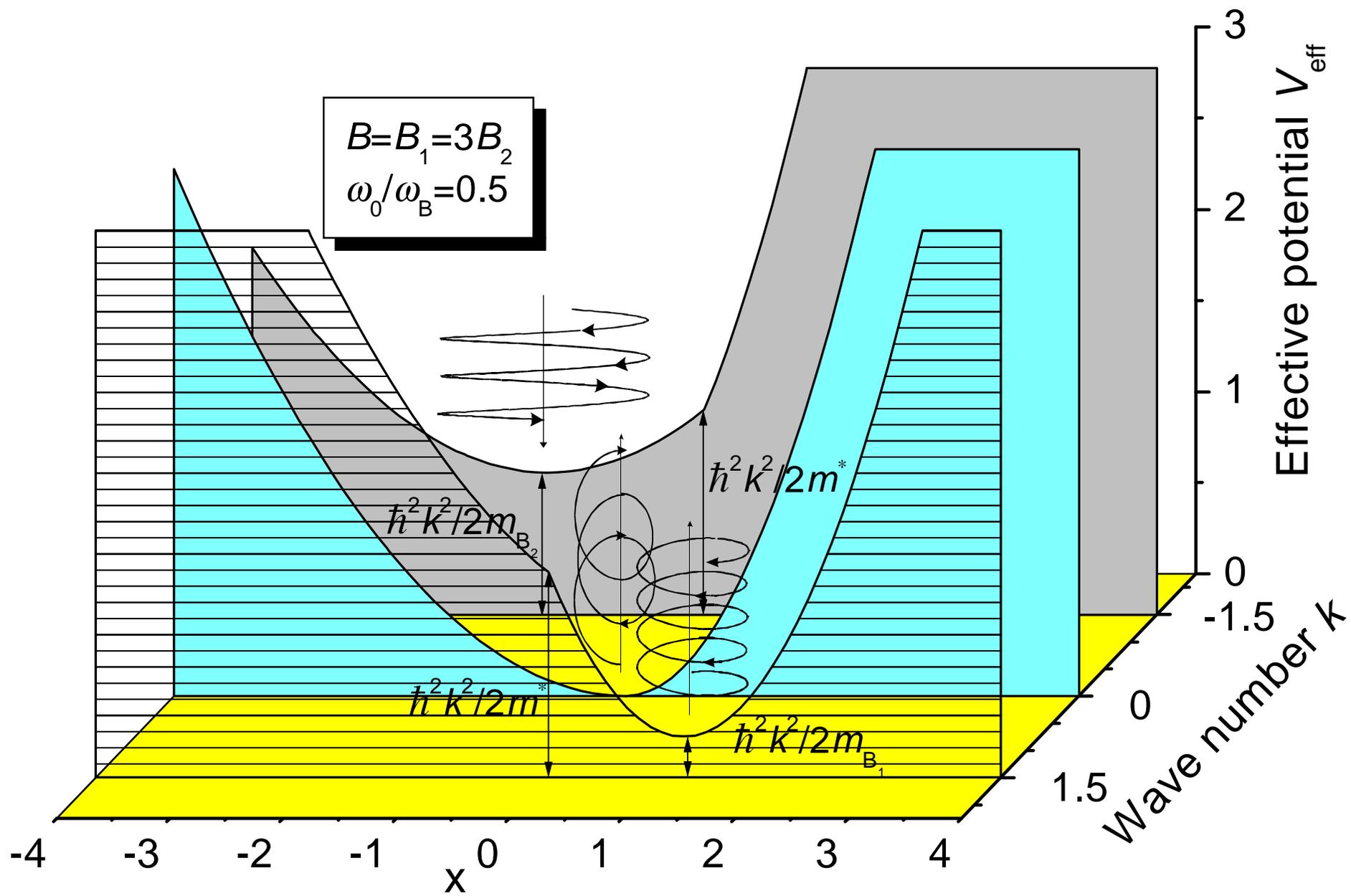
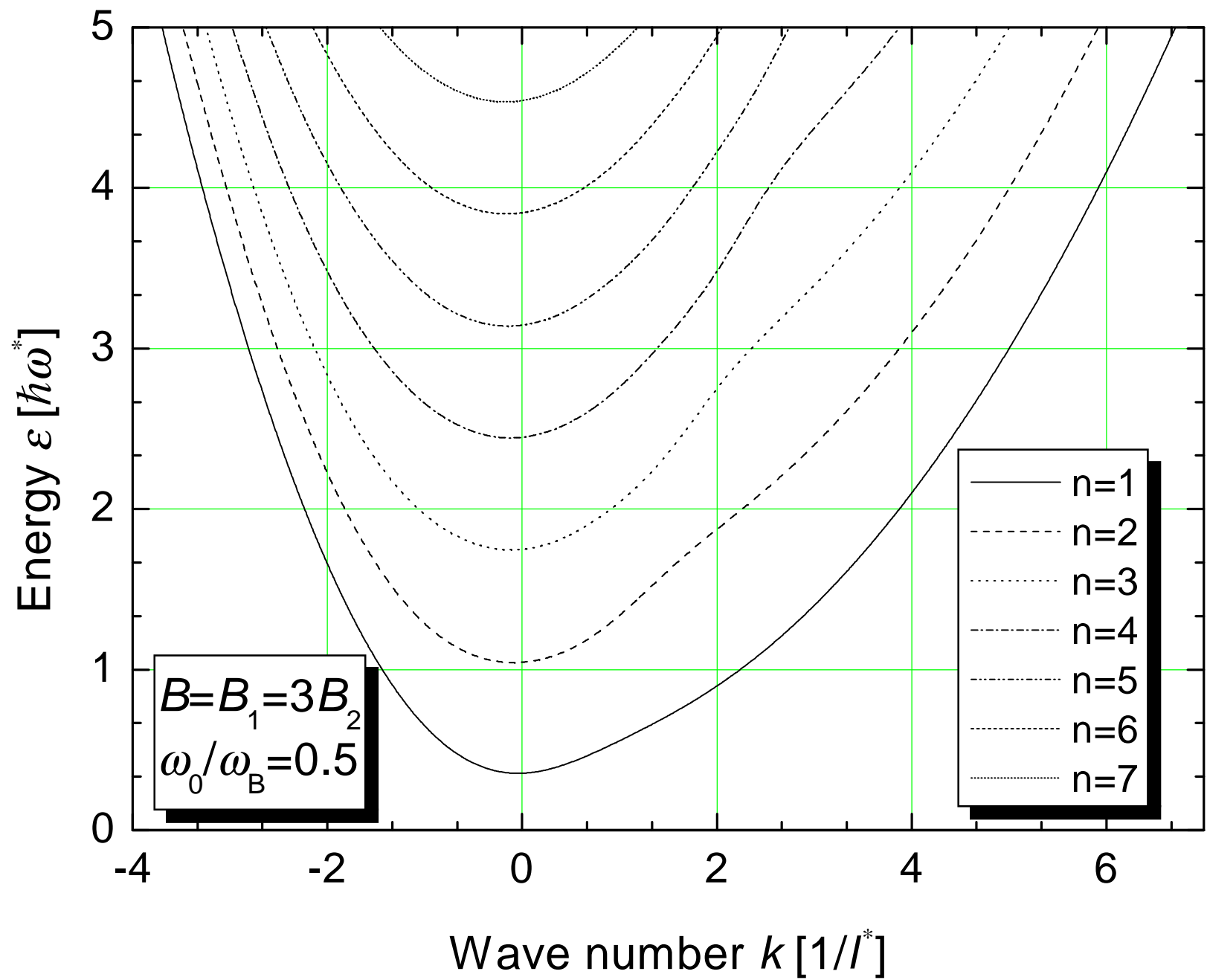
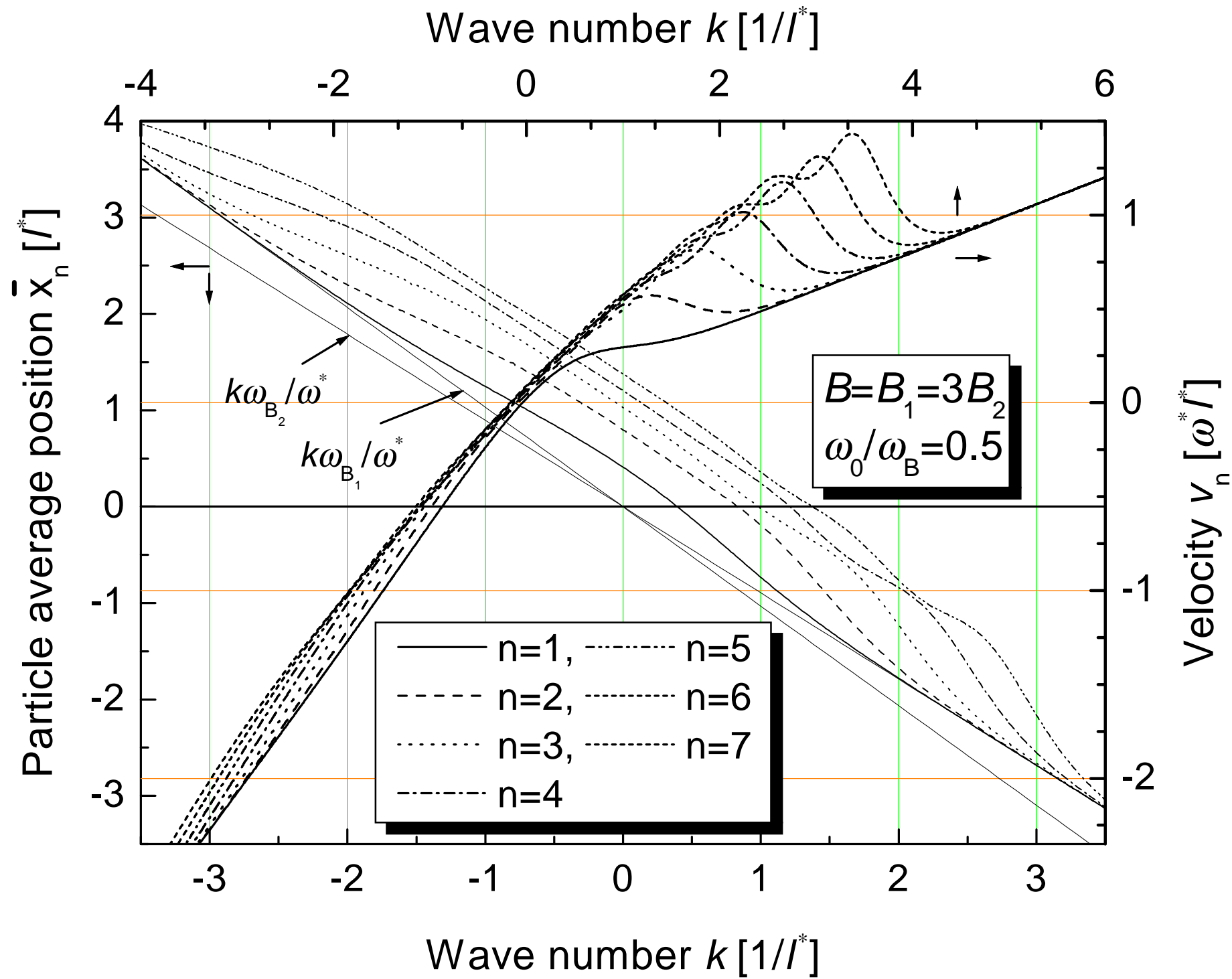
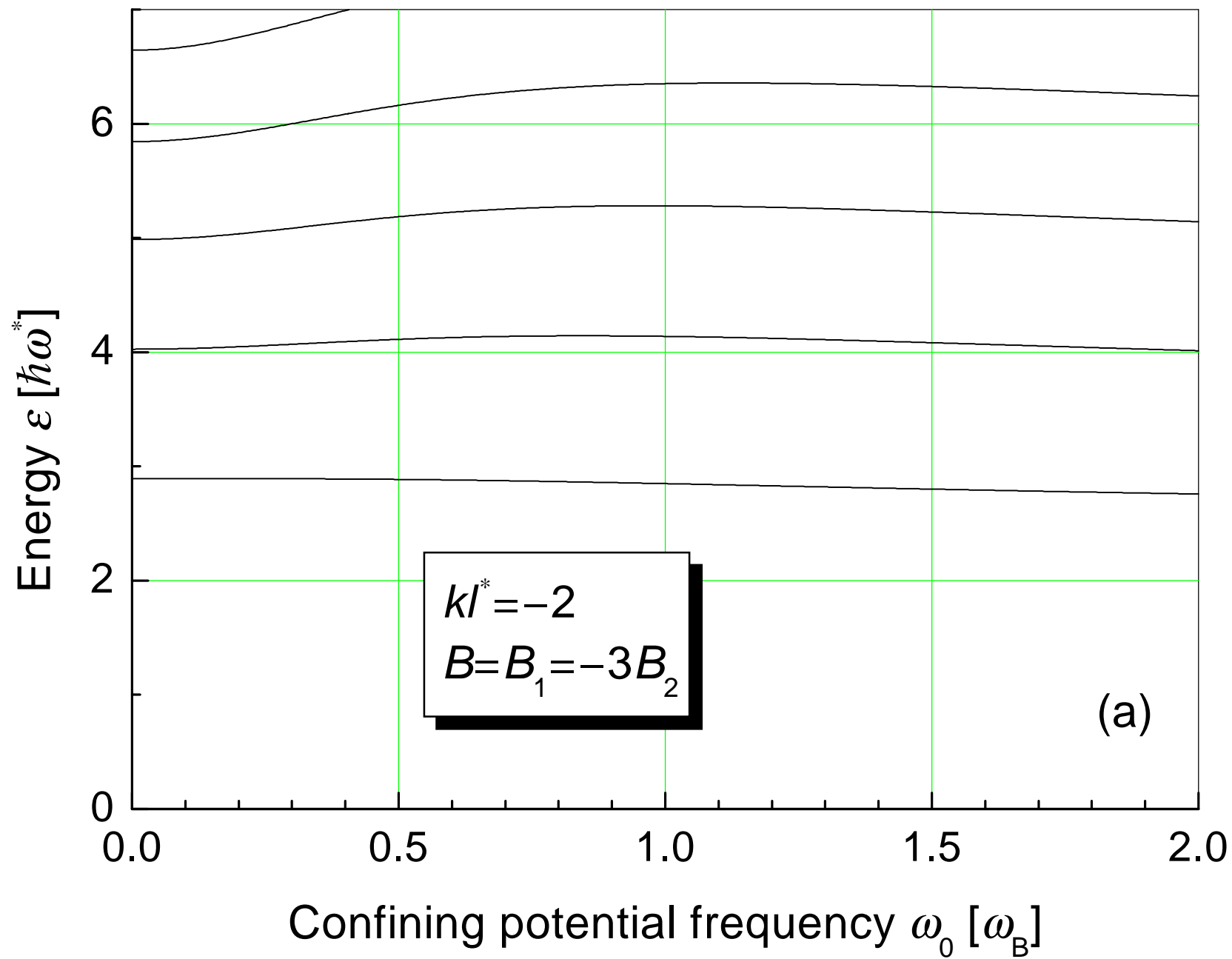
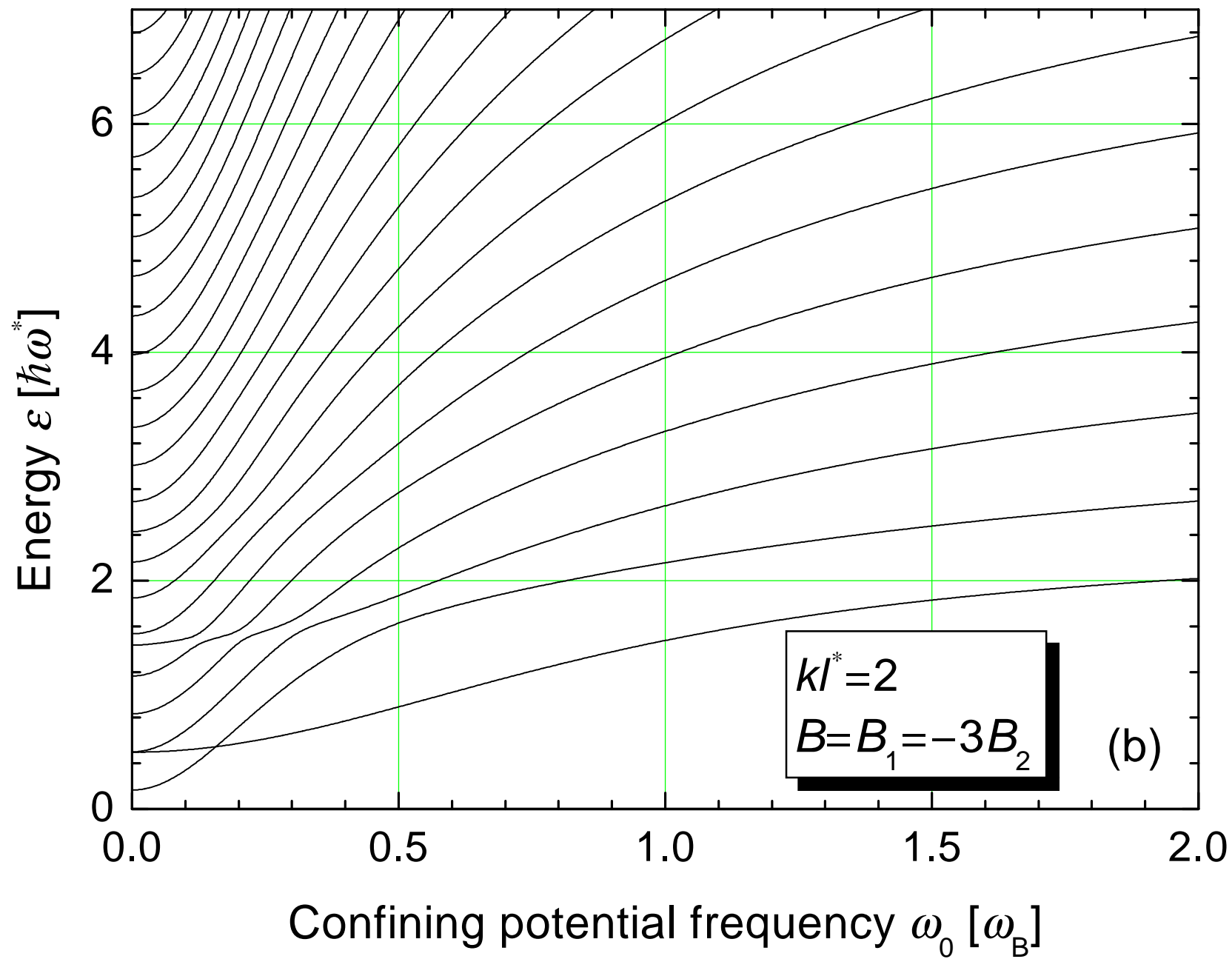
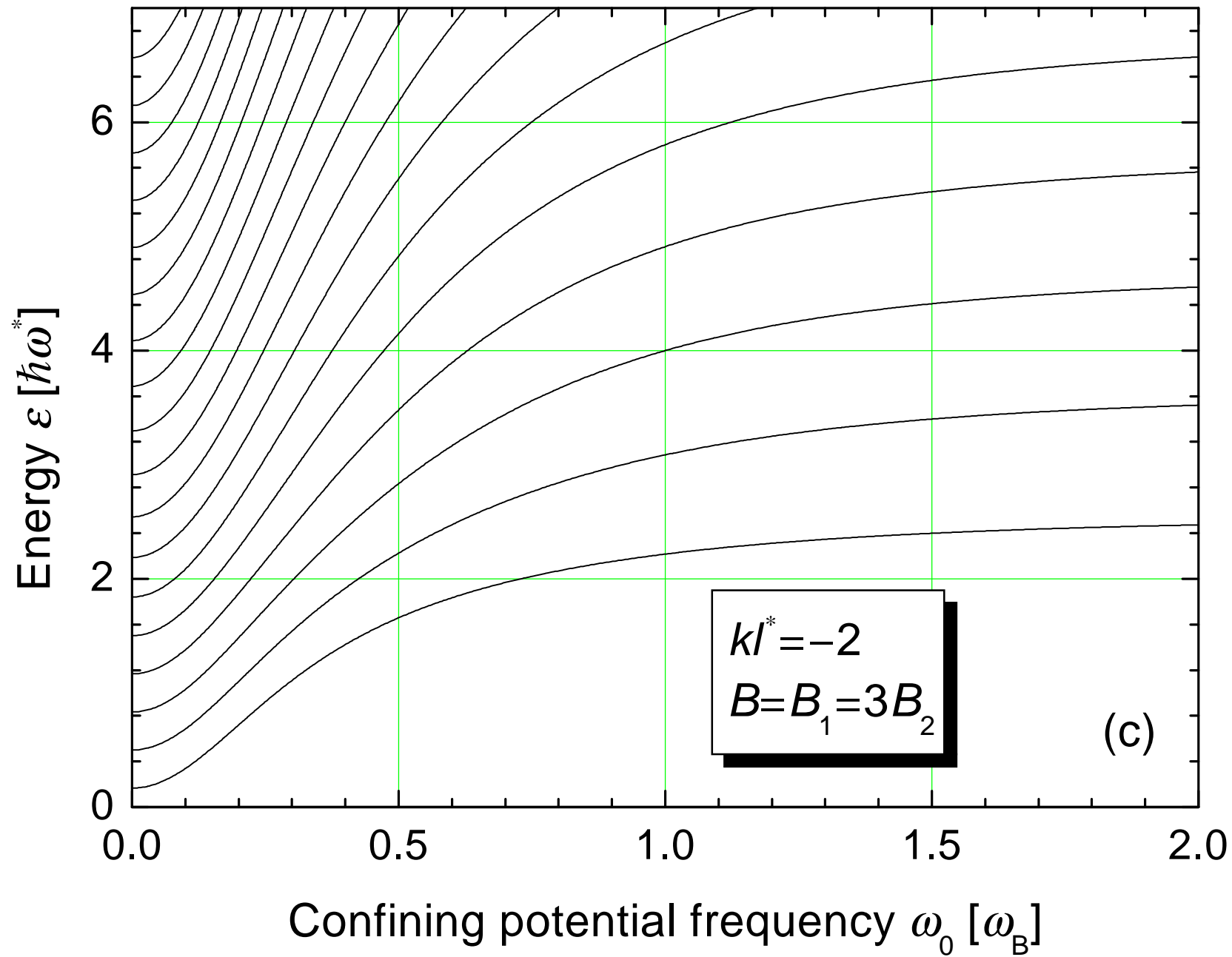
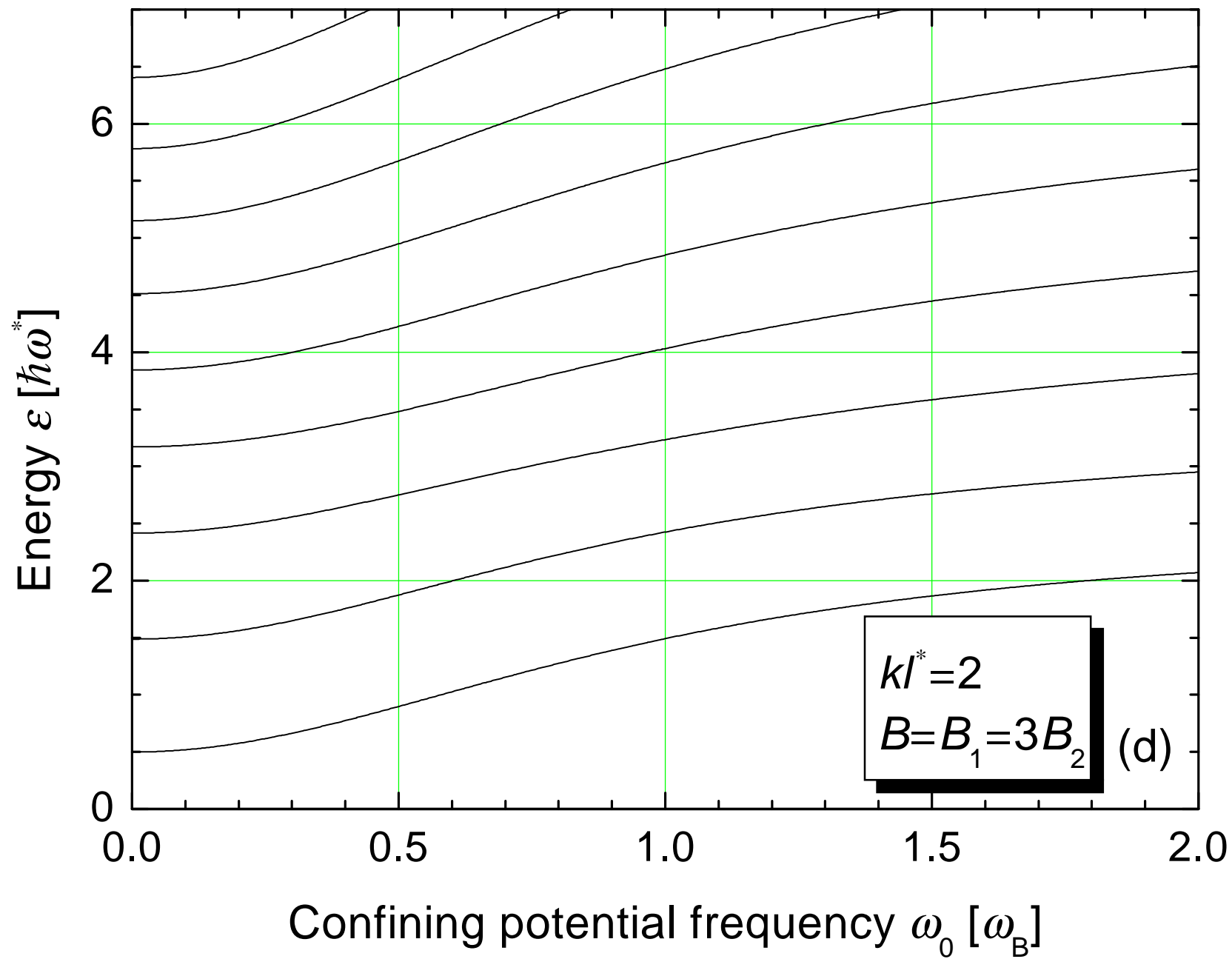
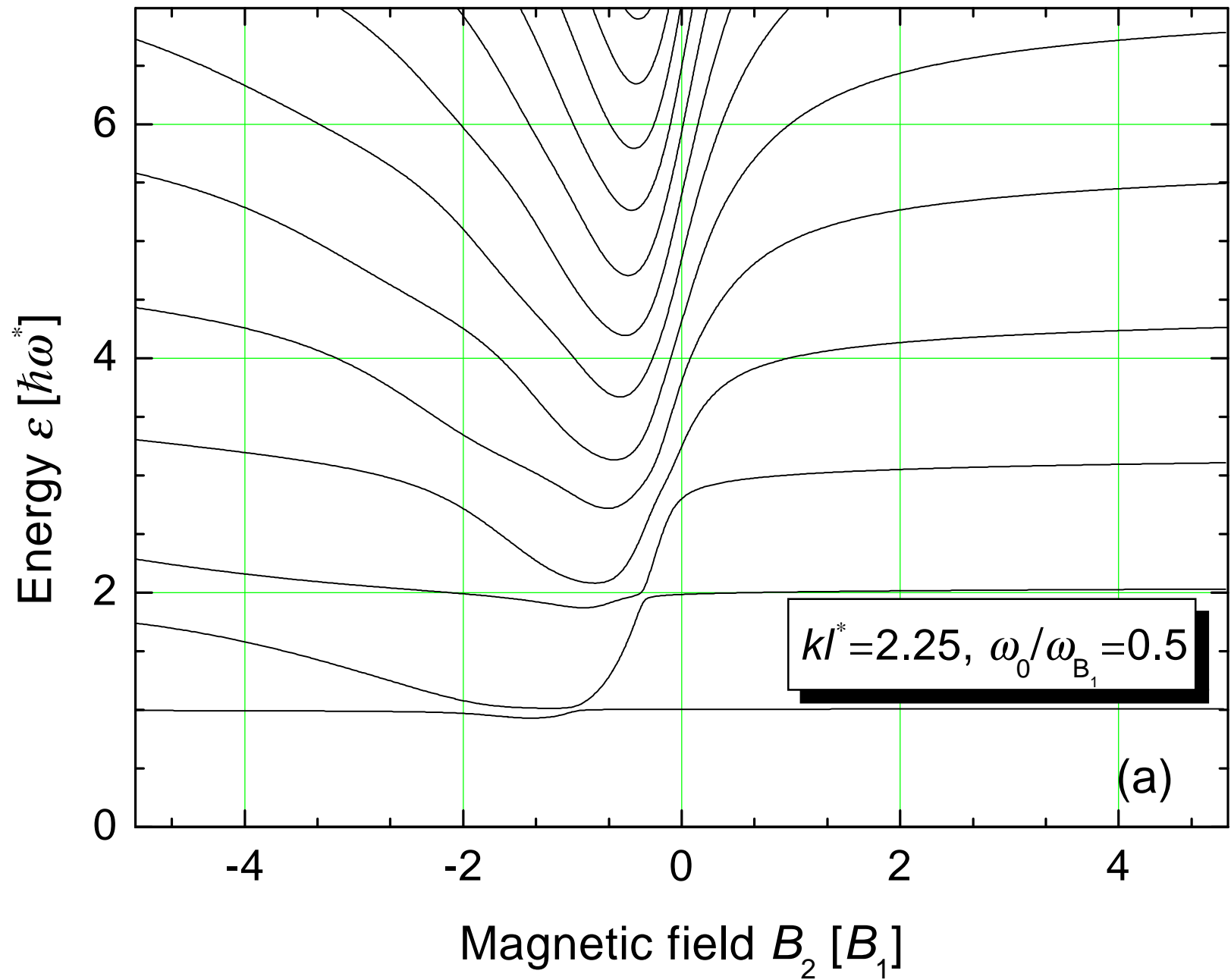
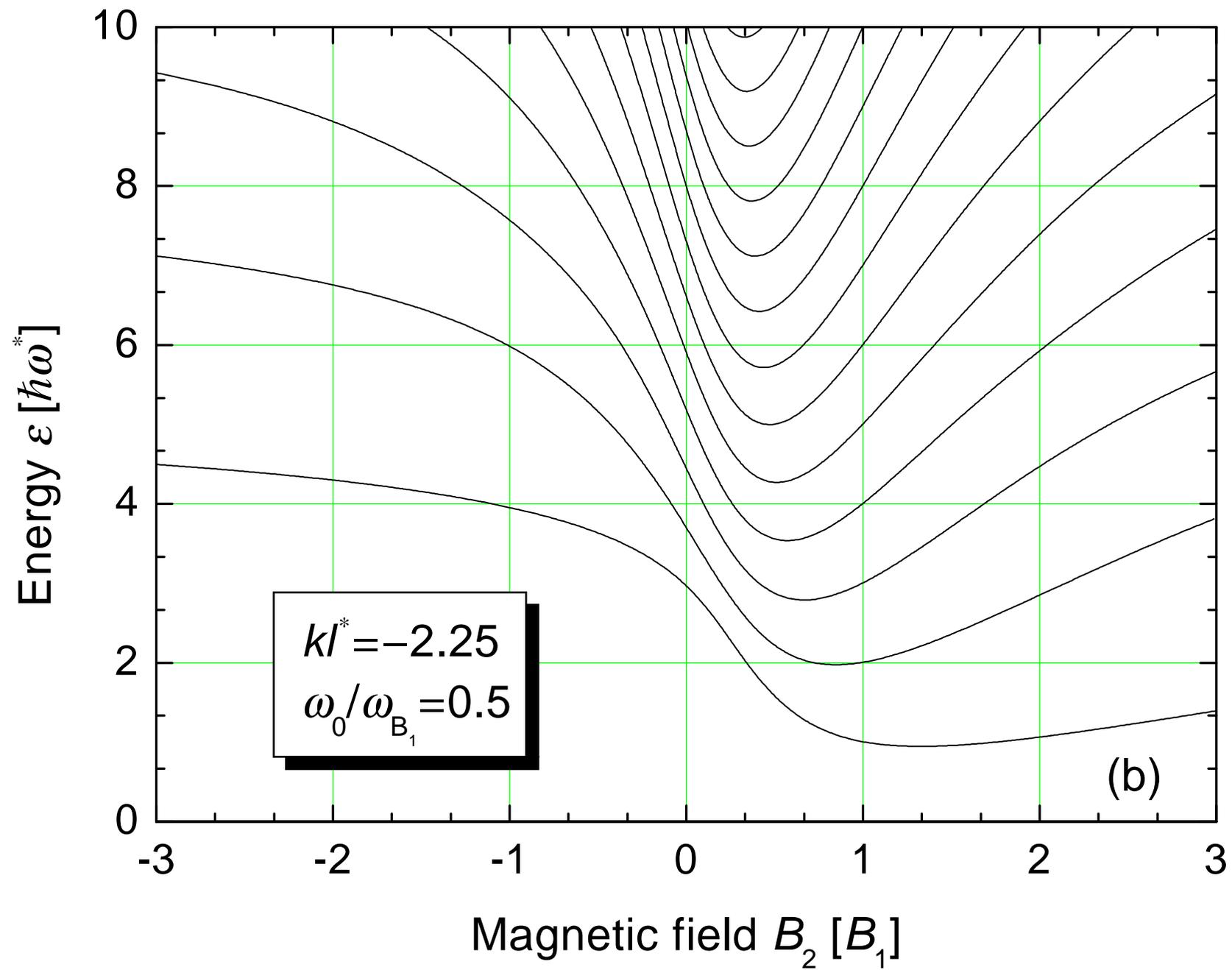
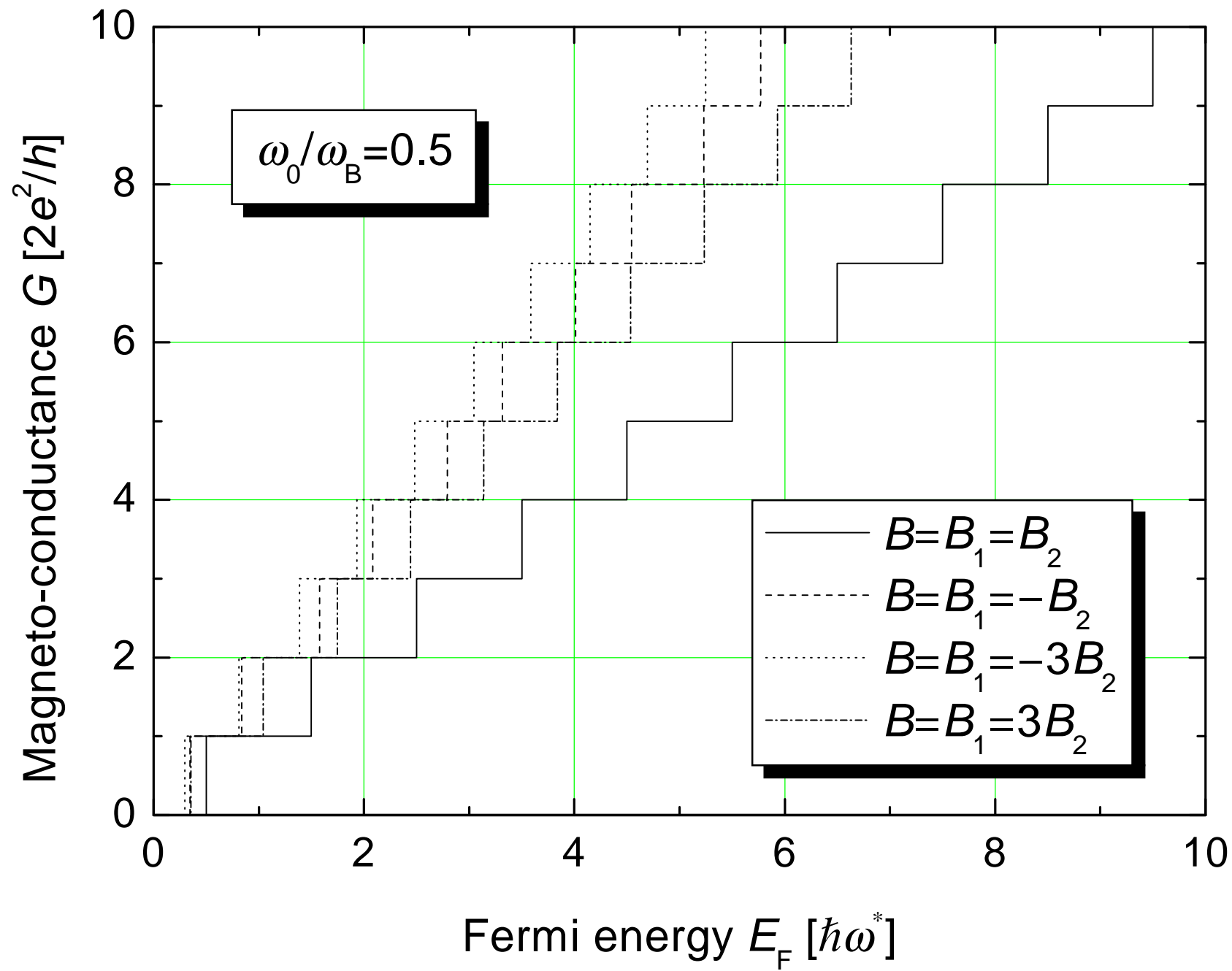
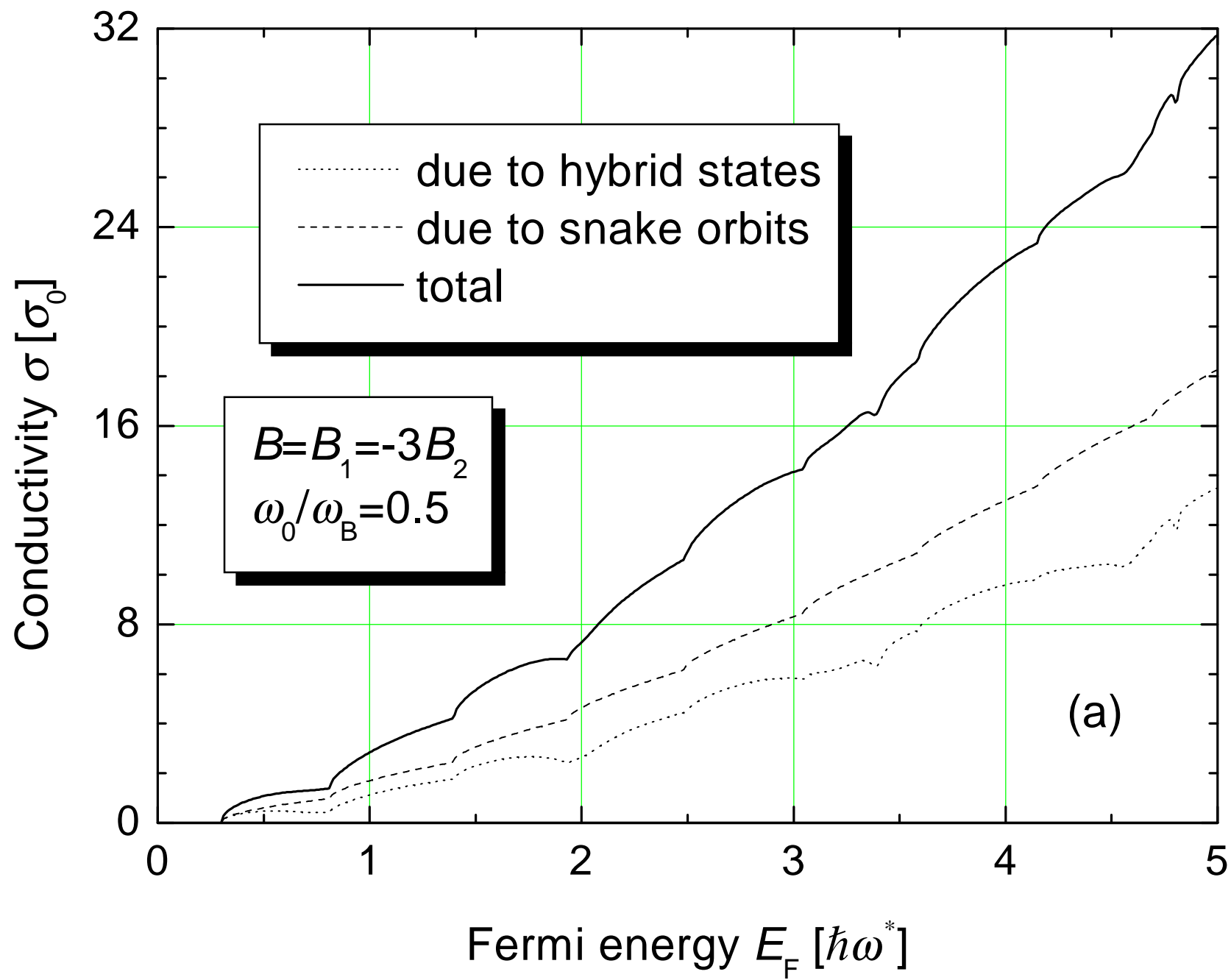
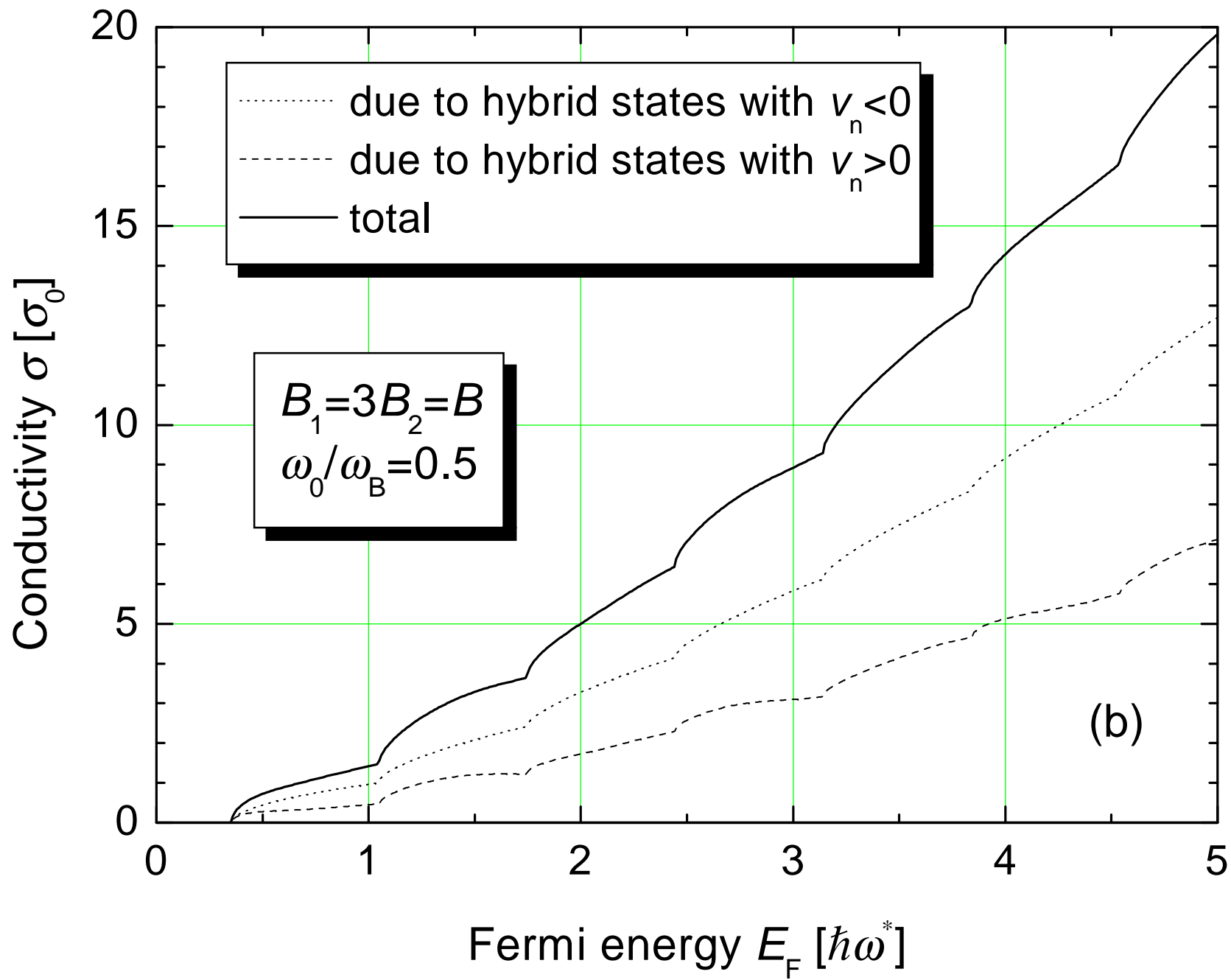
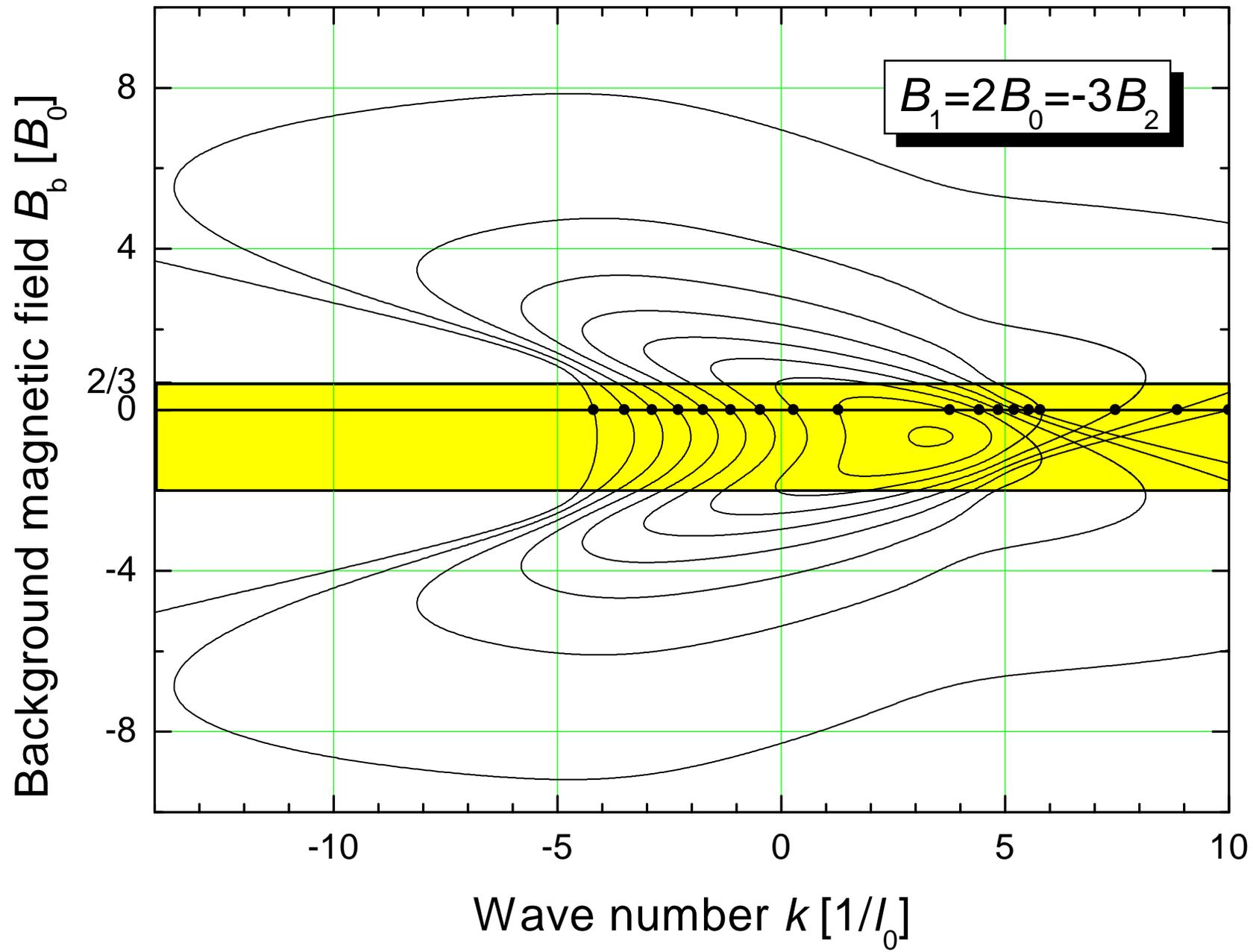
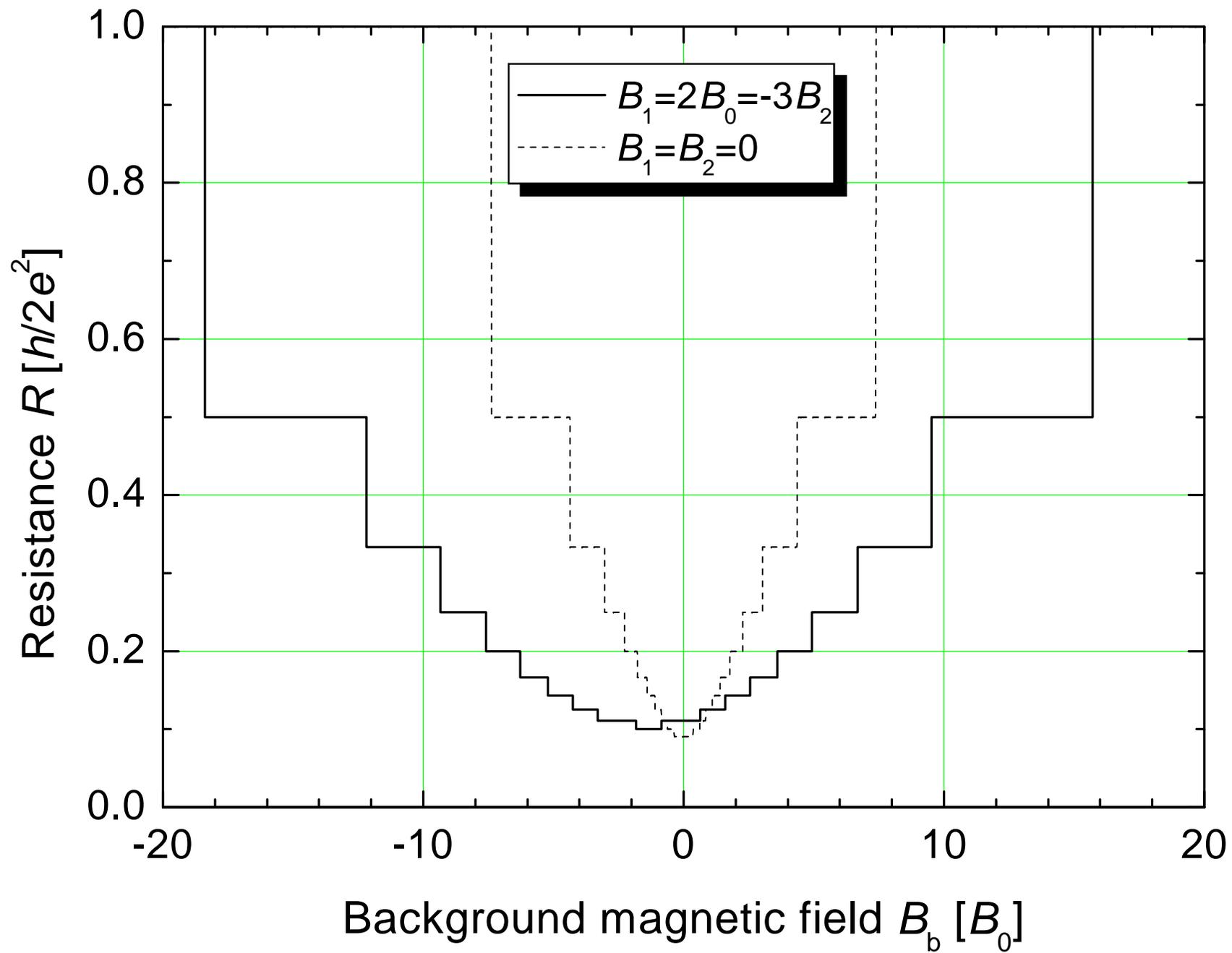
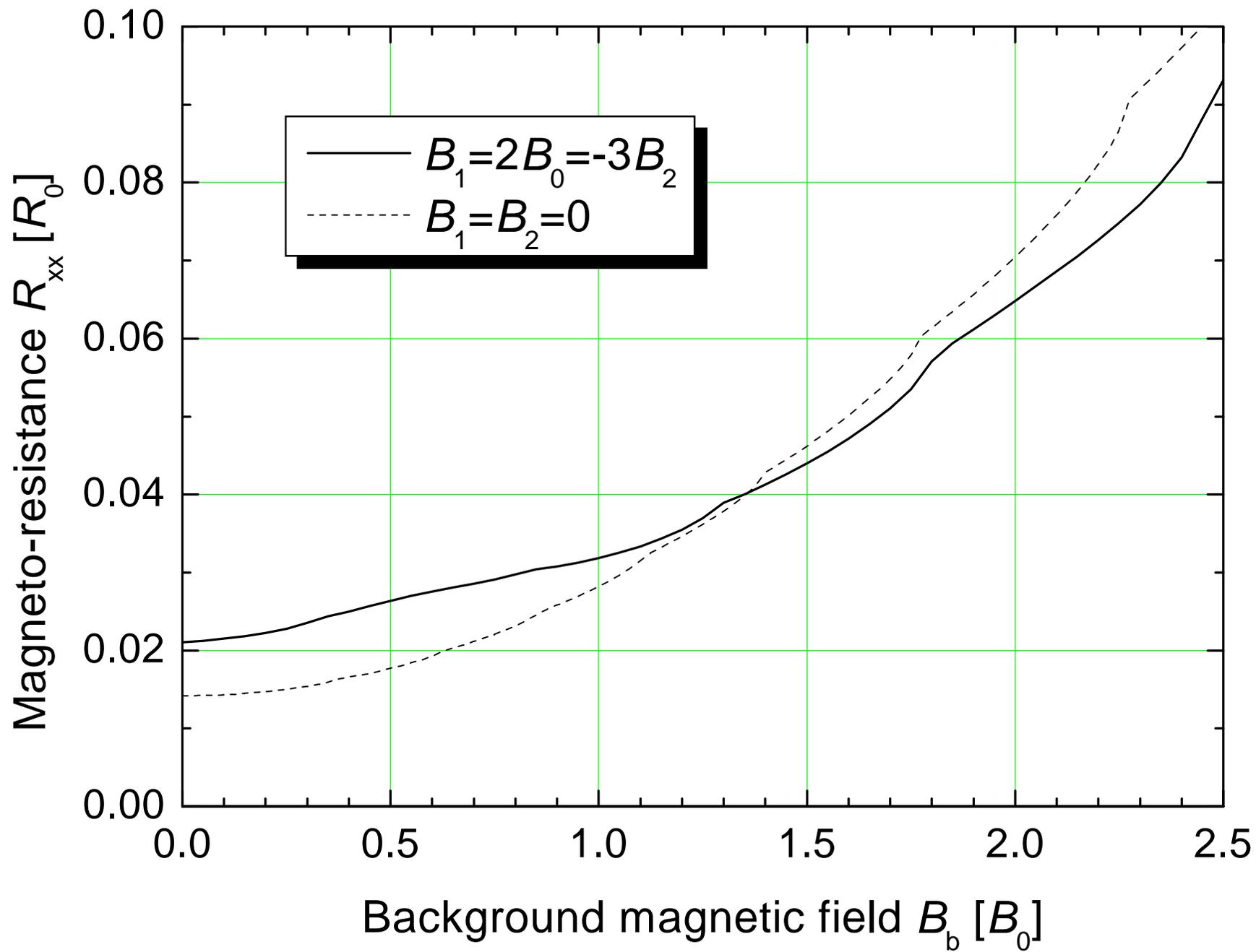
\end{document}